# Deep learning the sources of MJO predictability: a spectral view of learned features

Lin Yao*[1], Da Yang*[1], James P.C. Duncan[2], Ashesh Chattopadhyay[3], Pedram Hassanzadeh[1], Wahid Bhimji[4], and Bin Yu[5]

[1]University of Chicago, Chicago, IL, USA
[2]Allen Institute for Artificial Intelligence (AI2), Seattle, WA, USA
[3]University of California, Santa Cruz, Santa Cruz, CA, USA
[4]Lawrence Berkeley National Laboratory, Berkeley, CA, USA
[5]University of California, Berkeley, Berkeley, CA, USA

*Corresponding authors: Lin Yao, linyao@uchicago.edu; Da Yang, dayang@uchicago.edu

**Abstract**

The Madden-Julian oscillation (MJO) is a planetary-scale, intraseasonal tropical rainfall phenomenon crucial for global weather and climate; however, its dynamics and predictability remain poorly understood. Here, we leverage deep learning (DL) to investigate the sources of MJO predictability, motivated by a central difference in MJO theories: which spatial scales are essential for driving the MJO? We first develop a deep convolutional neural network (DCNN) to forecast the MJO indices (RMM and ROMI). Our model predicts RMM and ROMI up to 21 and 33 days, respectively, achieving skills comparable to leading subseasonal-to-seasonal models such as NCEP. To identify the spatial scales most relevant for MJO forecasting, we conduct spectral analysis of the latent feature space and find that large-scale patterns dominate the learned signals. Additional experiments show that models using only large-scale signals as the input have the same skills as those using all the scales, supporting the large-scale view of the MJO. Meanwhile, we find that small-scale signals remain informative: surprisingly, models using only small-scale input can still produce skillful forecasts up to 1–2 weeks ahead. We show that this is achieved by reconstructing the large-scale envelope of the small-scale activities, which aligns with the multi-scale view of the MJO. Altogether, our findings support that large-scale patterns—whether directly included or reconstructed—may be the primary source of MJO predictability.

## Introduction:

The Madden-Julian oscillation (MJO) is a planetary-scale envelope of individual rainstorms linked by propagating waves in the tropical atmosphere[1–5] (Fig. 1a). The MJO typically initiates over the Indian Ocean and propagates slowly eastward, with a typical life cycle lasting from 30 to 60 days—much longer than canonical weather systems. Since its discovery in the 1960s[1,2,6], studies have shown that the MJO modulates the frequency of tropical cyclone genesis[7–10], triggers El Niño events[3,11,12], and serves as a predictability source for sub-seasonal weather forecasts[5]. However, state-of-the-art climate models still struggle to accurately capture the MJO (e.g., rainfall variability, initiation, and propagation)[13–17], and the numerical weather prediction models have limited skills in predicting the MJO beyond 30 days[18] (about half of its period). These limitations arise because the primary source of MJO predictability remains uncertain, highlighting significant gaps in our understanding.

Current MJO theories diverge primarily on the role of spatial scales. One group of theories views the MJO as a large-scale wave[19,20], while the other emphasizes the critical role of small-scale convective activities embedded within the larger-scale envelope[21–24] (Fig. 1a). Both perspectives have been successful in explaining certain observed aspects of the MJO, but rigorous tests of their basic assumptions on the spatial scales remain difficult. For example, observational analyses often reveal correlations but fall short of establishing causality. On the other hand, mechanism-denial experiments in climate models provide a pathway to identifying causal relationships, but their results are often unreliable due to inaccurate representations of convection. This limitation highlights the need for a new approach to better study the MJO.

To close this gap, we use deep learning (DL) to investigate the primary spatial-scale sources of MJO predictability. Recent research has demonstrated that DL models can forecast the MJO 15 to 36 days in advance[25–28] (Table 1), with skills comparable to or even exceeding those of leading operational numerical models (20 to 30 days[29,30]). This promising development motivates us to explore the MJO predictability using a fully data-driven approach. To the best of our knowledge, previous DL work has primarily focused on improving forecasting skills, with limited emphasis on testing MJO theories or understanding the DL models themselves (Table 1). Here we emphasize understanding: we build a hierarchy of DL models that systematically vary input complexity (e.g., the number of variables and whether spectral filtering is applied) and model architecture. We conduct spectral analysis and sensitivity experiments to determine whether large- or small-scale patterns serve as the primary source of the MJO predictability in DL models.

## The DCNN model

In this section, we introduce a deep convolutional neural network (DCNN) to forecast the amplitude and location of the MJO, as characterized by the MJO indices. The most widely used index is the Real-time Multivariate MJO index[31] (RMM), a benchmark metric for inter-model comparison and operational monitoring. RMM has two components, RMM1 and RMM2, which are the first two principal components of anomalous outgoing longwave radiation (OLR, an indicator of deep convective clouds and precipitation systems), and zonal winds at 200 and 850 hPa, computed at the sub-seasonal timescale (Methods). RMM1 and RMM2 together identify the geographic location (phase) and amplitude (distance to the origin) of the MJO (Figs. 1b and S1). Another widely used index is the Real-time OLR MJO index[32] (ROMI; Figs. 1c and S1), which relies exclusively on temporally filtered OLR data. The major difference is that RMM captures variability based on both convection and circulations, while the ROMI only focuses on convection. Despite methodological differences, RMM and ROMI are highly correlated, and both effectively represent the amplitude and location of the MJO. Here, we report results for both to facilitate comparison across studies using different indices. Further details on RMM and ROMI are provided in Methods.

Fig. 2 gives an overview of our DCNN models. Our best model (BS) takes daily averaged snapshots of 18 variables over the tropical region as input (20°S - 20°N; Fig. 2a). These variables collectively represent the current state of the atmosphere and ocean surface. They include OLR, total column water vapor (TCWV), and sea surface temperature, along with thermodynamic and dynamic variables (humidity, winds, temperature, and geopotential) at three pressure levels corresponding to the lower, middle, and upper troposphere. The data preprocessing approach is described in Methods. Our model produces one-shot forecasts, meaning they are generated simultaneously rather than autoregressively. We implement two DCNN configurations: single-lead models (DCNN1; $n_{lead} = 1$), in which a separate network is trained for each prediction lead (0–35 days at 5-day intervals), and multi-lead models (DCNN2; $n_{lead} = 36$), in which a single network is trained to predict all leads simultaneously (0–35 days at daily intervals). To quantify the prediction uncertainty, we run 100 independent training realizations with identical data and architecture but different random initializations and data shuffling (hereafter ensemble members). Additional details on model architecture and training procedures are provided in Methods.

Both DCNN configurations show strong all-season MJO predictive skill, forecasting the RMM up to 21 days and the ROMI up to 33 days in advance using the full 18-variable input set (see Fig. S2 for DCNN1 and Fig. 2c–d for DCNN2). Prediction skill is quantified by the bivariate correlation coefficient (BCC) between the forecasts and truth, using a 0.5 threshold denoting useful skill (Methods). In Fig. 2c–d, the colored lines indicate the prediction skill of the ensemble mean: we first average predictions across all ensemble members and then compute the BCC between this mean and the truth. The shading represents model uncertainty. At longer lead times, the models fail to fully capture variability at the tails of the observed MJO index distributions (missing extremes), leading to reduced performance (Figs. S3–8). Beyond the general skill measured by BCC, the evolution of the MJO is visualized in phase diagrams for RMM and ROMI (Figs. 1b–c and S1). Each curve represents the averaged trajectory of MJO events starting from a specific phase. The single-lead model (DCNN1) performs slightly better than the multi-lead model (DCNN2) at shorter leads, while their performance converges at longer leads, leading to the same prediction skills. For the RMM, both location and amplitude are well predicted, whereas for the ROMI, the location is well predicted, but the amplitude is substantially underestimated. The BCC is more sensitive to the location than the amplitude, which explains the longer predictive skill obtained for ROMI. Increasing architecture complexity slightly (Figs. S9–10) or using alternative OLR datasets (NOAA vs. ERA5) does not significantly influence the prediction skill. Additional metrics, including root mean square error and skill dependence on the initial phases and seasons, are provided in the Supplementary Information (Figs. S11-13).

For winter MJO, our models achieve predictive skill up to 25 days for RMM (Fig. S14). The performance of our DCNN model is comparable to most numerical models in the Subseasonal-to-Seasonal (S2S) prediction project[29,30], such as those from the National Centers for Environmental Prediction (NCEP), the UK Met Office (UKMO), and Météo-France/Centre National de Recherches Météorologiques (CNRM). Only the European Centre for Medium-Range Weather Forecasts (ECMWF) demonstrates significantly superior skill, with accurate RMM predictions extending to 30 days. Compared to previously developed DL-based MJO forecasting models (Table 1), our DCNN models match or exceed most existing results[25-27], except for FuXi-S2S[28], an autoregressive model that relies on a substantially more complex encoder-perturbation-decoder architecture and 76 input variables. Thus, our DCNN models may provide a promising data-driven approach that complements numerical modeling to advance understanding of the MJO.

In the following sections, we construct two types of DL model hierarchies. The first varies the input complexity to identify the primary sources of MJO predictability. The second simplifies the model architecture by evaluating the contribution of each layer, reducing the original U-Net to a shallower yet effective structure. For convenience, we will focus on the multi-lead model (DCNN2) using NOAA OLR in the main text from this point forward. Results using other configurations—including single-lead models and models using ERA5 OLR—are explicitly noted when used and are provided in the Supplementary Information.

**Source of Predictability**

To investigate MJO predictability, we construct DL models with varying levels of input complexity (Fig. 2a and Table 2). By progressively simplifying the input, we aim to identify what spatial scales matter more for predicting the MJO indices. This might provide DL-based evidence for the two schools of MJO theories discussed in the introduction.

First, we simplify the input by using a single variable instead of the full set of 18 variables. Despite this substantial reduction in input complexity, the one-variable model still provides skillful prediction of RMM up to 15 days and ROMI up to 26 days in advance (green lines in Fig. 2c-d). Across the 18 candidates, TCWV and OLR yield the highest prediction skill for RMM and ROMI, respectively (Fig. S15). Note that a variable that gives the highest forecast skill need not be the dynamical cause of predictability; it may be a correlated proxy. The "best" input variable can depend on the data source, preprocessing pipeline, model architecture, and training configuration[27]. For example, in our single-lead model using ERA5 OLR, OLR performs as well as TCWV for predicting RMM (Fig. S16). Moreover, the 18 variables are highly correlated, so omitting other variables may not remove their combined impacts. Accordingly, we use the one-variable models (CT) to reduce complexity and to provide a control for the later filtered-input experiments (Table 2), rather than to infer physical causation.

Before filtering the inputs, we examine the spectra of the inputs and feature maps in CT models. Matsuno (1966) developed a theory for equatorial waves using linear shallow water equations on an equatorial $\beta$-plane. This theory shows that the basis functions of the equatorial waves in the zonal and meridional directions are the Fourier series and the parabolic cylinder functions, respectively. Therefore, we project the inputs and convolutional feature maps as follows:

$$\hat{q}(k,m) = \int_{-y_0}^{y_0} \int_0^{2\pi R} q(x,y) \cdot e^{-\frac{ikx}{R}} \phi_m\left(\frac{y}{L_d}\right) dx\, d\left(\frac{y}{L_d}\right), \quad (1)$$

where

$$\phi_m(\xi) = \left(m!\sqrt{\pi}2^m\right)^{-\frac{1}{2}} e^{-\frac{\xi^2}{2}} H_m(\xi).$$

Here, $q$ represents TCWV, shown as an example of the maps we project. $k$ and $m$ are integers, representing zonal and meridional wavenumbers. $x$ and $y$ represent the distances in the zonal and meridional directions, and $R$ represents Earth's radius. $y_0$ is the distance from the equator to the northern boundary of the domain. $L_d$ is the equatorial deformation radius, given by $\sqrt{c/\beta}$, where $c$ is the gravity wave speed, and $\beta$ is the meridional gradient of the Coriolis parameter. $H_m(\xi)$ is the $m^{th}$ Hermite polynomial. The Fourier series and $\phi_m(\xi)$ thus provide orthogonal bases that transform latitude-longitude maps into $(k, m)$ spectral space (Methods). We restrict our analysis to feature maps in the last convolutional layer (L2+6 in Fig. 2b) that make essential contributions to the model output. Spectral analysis has been shown to effectively reveal how convolutional layers filter physically relevant signals in turbulence and climate modeling[34,35]. Building on this, we analyze the power spectra of inputs and feature maps to assess which spatial scales are primarily transmitted through the convolutional layers for forecasting.

Spectral analysis reveals that large-scale signals dominate the input and are preferentially transmitted through the network for MJO forecasting. Fig. 3a–b shows the log-scaled power spectra for CT models predicting RMM. In both the input and final-layer feature maps, most power is concentrated at large scales—corresponding to zonal and meridional wavenumbers below 10 (> 4000 km zonally). Relative to the input, the feature maps exhibit amplified power at the largest scales (e.g., zonal wavenumbers 1 and 3; Fig. S17), reduced power in the intermediate range (wavenumbers 4–10), and a modest increase at smaller scales (wavenumbers >10). A similar pattern is observed for CT models predicting ROMI, with consistent amplification of large-scale power, particularly at zonal wavenumbers below 10 (Figs. 3c and S17). Collectively, the results indicate that large-scale features are preferentially retained and enhanced by the network, underscoring their important role in forecasting the MJO across both RMM and ROMI targets.

To test whether large-scale signals are essential for MJO forecasting, we further simplify the input to spatially filtered variables (LA-SS in Fig. 2a). As summarized in Table 2, we first construct large-scale experiments by removing small-scale components—initially in the meridional direction (LA), and then in both zonal and meridional directions (LL). The cutoff wavenumber is set to 10 based on our spectral analysis, corresponding to a zonal (meridional) scale of approximately 4000 (1000) km. We also construct a small-scale experiment (SS), where only signals above this cutoff are retained in both directions. Because each wave mode is orthogonal to the others, different scales are linearly independent from each other. Such filtering is closely analogous to disabling large- or small-scale processes in numerical models (i.e., mechanism-denial experiments).

We find that the large-scale experiments (LA and LL) retain most of the prediction skill observed in the control experiments (CT), while the small-scale experiments (SS) show a substantial reduction in model performance (Fig. 2c–d). We design two configurations: (1) applying filtered inputs directly to the pretrained CT models without retraining (dashed lines), and (2) retraining new DCNNs from scratch (i.e., random weights) using the filtered inputs (solid lines). In both cases, LA and LL models achieve comparable performance to CT, predicting RMM up to 14 days and ROMI up to 25 days in advance, with only minor degradation in skill. In contrast, SS models consistently underperform. When applied without retraining, they fail to produce meaningful forecasts even at day 0; when retrained, they can only predict RMM and ROMI up to 8 and 9 days, respectively. The skill can be slightly improved by using single-lead models (DCNN1)—up to 9 days for RMM and 14 days for ROMI—owing to better shorter-lead performance compared to multi-lead models. All experiments use tuned hyperparameters selected from the best-performing models (Methods). The results remain robust across a range of settings: when the input domain is expanded from 20°S–20°N to 40°S–40°N, when basis functions are based on different gravity wave speeds, when small-scale experiments are defined differently, and when other input variables are used (Supplementary Text 1 and Figs. S18–21), further validating the conclusions drawn from SS. Altogether, these results are consistent with the spectral analysis and highlight that large-scale patterns are more effective in MJO forecasting, consistent with the large-scale theories[19,20].

However, the role of large-scale patterns is sensitive to the choice of cutoff wavenumbers (Supplementary Text 2 and Figs. S22–23). In the main text, we use a cutoff wavenumber of 10 in both zonal and meridional directions, based on spectral analysis. However, large-scale MJO theories typically assume the MJO is of wavenumbers 1 to 3 (planetary scale). As the cutoff wavenumbers decrease (e.g., from 10 to 3), the prediction skill of the large-scale experiments (LL) consistently declines, while the skill of the small-scale experiments (SS) increases. This means that processes between 4000 km and 13000 km make a meaningful contribution to MJO forecasts, indicating multi-scale interactions. This trend holds in both multi-lead and single-lead models. Notably, in single-lead models, when the zonal cutoff is reduced to five (~8000 km) and the meridional cutoff to three—a generous threshold for distinguishing large- and small-scale processes in MJO theories—the prediction skill of the large-scale experiments can fall below that of the small-scale experiments (Fig. S24). This indicates that small-scale processes (<8000 km) might be as critical as large-scale ones in driving MJO dynamics. This seems to align with the multi-scale theories[21–24].

To further investigate the role of small-scale signals, we analyze the retrained SS models in Fig. 2 (using a cutoff wavenumber of 10), which are still able to produce reasonable MJO forecasts up to 1–2 weeks in advance. Spectral analysis of final-layer feature maps reveals that the convolutional layers can, surprisingly, reconstruct large-scale patterns from small-scale inputs, and use these recovered signals to make MJO forecasts (Fig. 3c, d, g, and h). Again, spectral analysis is restricted to essential feature maps (Methods). Most of the power after convolution is concentrated at wavenumbers below 10, indicating the dominance of large-scale signals. A layer-by-layer analysis further shows a progressive amplification of large-scale patterns throughout the network (Fig. S25). These results are consistent across multi-lead and single-lead models. Together, the spectral analyses on SS models suggest a transition from small- to large-scale representations within the network, highlighting the essential role of large scales in making the final one-shot forecasts of the MJO.

To understand how DCNN models reconstruct large-scale patterns, we examine individual input fields and final-layer feature maps from the SS experiments. Fig. 4 presents an example from December 15, 2021. Although large-scale information is removed from input, the MJO convective envelope remains visible through the spatial organization of small-scale structures (Fig. 4a, b, e, and f). We identify the top ten most essential feature maps from the final convolutional layer (Fig. S26)—based on their contributions to the output[36]—and compute a feature map composite (Methods). The resulting maps (Fig. 4c, d, g, and h) show clear wave packets for models predicting RMM and ROMI, such as the one extending from 30 °E to 150 °W. Additional examples are shown in Fig. S27, exhibiting similar patterns. These results suggest that the DCNN reconstructs large-scale envelopes from small-scale input via successive convolution and nonlinear activation. The model's reliance on recovered large-scale patterns, rather than direct use of small-scale features, supports the conclusion that large-scale signals are more effective in making the one-shot MJO forecasts. This particular reconstruction process seems to align with multi-scale theories that view the MJO as a large-scale envelope of small-scale convective waves[21–24].

**A simpler DL architecture**

Apart from simplifying the input, can we build a simpler model using a simpler architecture that favors interpretability[37]?

To answer the question, we first evaluate the contribution of different convolutional layers. Following Shin et al. (2022), we analyze the sensitivity of the CT model outputs to changes in feature maps in the convolutional layer. To test the robustness of our findings, we apply two complementary approaches: zeroing out feature maps and adding small perturbations. In the first method, we sequentially zero out individual feature maps in the last convolutional layer (L2+6) and compare the modified output to the original (Methods). As shown in Fig. S28, only a subset of feature maps significantly affects the RMM or ROMI output, with the most influential ones originating from the shallower layer L2. These results hold across both multi-lead and single-lead models at various lead times (Figs. S28–29).

To validate the zeroing-out method, we conduct the second method by adding small perturbations to individual feature maps (Methods) and evaluating the output sensitivity. Due to the high computational cost, this method is applied only to single-lead models using ERA5 OLR as input. The results from both methods are consistent, reaffirming the major contribution of feature maps in L2 (Fig. S30). We also find that even when we zero out all feature maps in deeper layers (L3 to L6) in the trained-CT model, the model still provides the same level of prediction skill as the original one without retraining (Fig. S31). This result holds in both multi-lead and single-lead models. These findings motivate a reduction in model depth.

We therefore simplify our DCNN to a shallow CNN with only the first two convolutional layers (L1 and L2; Fig. S32a), removing deeper layers (L3 to L6) that contribute little to the output. The CNN is trained from scratch and achieves prediction skills equivalent to the original CT DCNN models. It forecasts RMM up to 15 days and ROMI up to 25 days in advance (Fig. S32b), while requiring only one-third of the trainable parameters. We then perform filtered-input experiments using the shallow CNN in single-lead configurations. Models using large-scale input signals outperform those using small scales, consistent with results from DCNN-based experiments (Fig. S33). This might suggest that our findings do not depend on specific model architectures, and that similar filtered-input experiments can be applied to other DL models.

**Summary and discussion**

We employ DL models to investigate the primary source of MJO predictability, motivated by a fundamental question in MJO theory: which spatial scales are most essential for forecasting the MJO? Our DCNN models predict RMM and ROMI up to 21 and 33 days in advance, respectively, comparable to many numerical models in the S2S prediction project. Our models are also the best

DL models among one-shot forecasting methods. The good performance of our DL models enables us to test hypotheses grounded in MJO theories using a fully data-driven approach.

To examine the contribution of different spatial scales, we construct a DL model hierarchy with three levels of input complexity. We begin with models using the full set of 18 tropical variables (BS) and then simplify the input to a single variable (CT). We find that TCWV yields the highest skill for RMM and OLR for ROMI, achieving skillful forecasts up to 15 and 26 days, respectively. We then analyze the feature maps from the final convolutional layer of CT models using spectral analysis based on equatorial wave theories (Matsuno, 1966). This analysis reveals that large-scale patterns dominate model latent representation and are preferentially transmitted to the output layers.

To validate the analysis, we further reduce the input complexity by filtering out either large-scale or small-scale patterns from the input. This approach removes the linear correlation between large- and small-scale patterns, analogous to mechanism-denial experiments in numerical models[38-40]. The cutoff wavenumber is set to 10, corresponding to a zonal scale of ~4000 km. We find that models using only large-scale information as the input (LA and LL) have the same prediction skill as CT, while the small-scale experiments (SS) significantly lose prediction skills. These results demonstrate that large-scale signals are more effective in making the one-shot forecasts of the MJO, consistent with large-scale MJO theories[19,20].

At the same time, we find compelling evidence supporting multi-scale MJO theories[21–24]. First, despite the removal of large-scale information (>4000 km zonally), SS models still produce skillful forecasts up to 1–2 weeks ahead. Analysis of input and feature maps shows that the model reconstructs the large-scale envelope from small-scale inputs through successive convolution and nonlinear activation (Fig. S25). Second, the advantage of large-scale models depends on the choice of spectral cutoff. When the cutoff wavenumbers are reduced from 10 to 3 in both directions (~13000 km), the small-scale model (SS) outperforms the large-scale model (LL; Fig. S24). This finding highlights the contribution of processes between 4000 km and 13000 km to predict the MJO, indicating potential multi-scale interactions.

Together, our results provide data-driven evidence supporting both large-scale and multi-scale views of the MJO. Previous studies have extensively demonstrated observational evidence in favor of large-scale theories, which highlights the positive feedback between moisture and convection[41] and the crucial role of planetary-scale waves[42] in driving MJO dynamics. However, observational support for multi-scale theories has been comparatively scarce. While our study does not definitively determine which theory is correct, it offers independent, observation-based evidence that supports the role of small-scale processes and thus multi-scale theories[21–24]. More broadly, this work represents a step toward explainable AI for climate, achieved by combining physics, spectral analysis, and ML techniques. Such efforts remain rare but are increasingly recognized as essential, as emphasized by Lai et al. (2025).

In addition to hypothesis testing, we apply explainable AI tools to simplify model architecture. We find that only a subset of feature maps, primarily from L2, contributes to the output (Figs. S28-30). Zeroing out all feature maps from L3 to L6 in the trained model does not affect prediction skill (Fig. S31), leading us to simplify the U-Net into a much simpler CNN with only two convolutional layers (Fig. S32a). This simplified CNN achieves the same prediction skill as the original DCNN. These findings help explain why previous CNN-based studies[26,27] achieved a similar performance to our DCNN model. They also underscore the value of critically evaluating and removing unnecessary components in DL models to improve efficiency and interpretability.

Our results may depend on how the MJO is represented. In this study, we rely on the RMM and ROMI indices, both derived from large-scale fields. While these indices are widely used, they are simplified, low-dimensional representations of the MJO. As such, the conclusions drawn from our model may be influenced by how well these indices capture the full complexity of the phenomenon. Important information relevant to the sources of MJO predictability may be absent from these indices. To further validate our findings, we plan to examine other indices of the MJO (e.g., the large-scale precipitation tracking method[44,45]). It is also desirable to develop an alternative MJO index that avoids spatial filtering and to revisit our findings in that framework.

| Paper | Hypothesis testing | Input | Architecture | Performance for winter MJO | Performance for all-season MJO |
|---|---|---|---|---|---|
| Martin et al. (2021) | **NA** | 3 variables | Fully connected network (FCN) | 18 days | / |
| Delaunay and Christensen (2022) | **NA** | 7 variables | CNN + FCN | / | 20 days |
| Shin et al. (2024) | **NA** | 5 variables with transfer learning | CNN + FCN | 25 days | / |
| This paper | **YES** | 18 variables | CNN + FCN | 25 days | 21 days |
| Chen et al. (2024) | **NA** | 76 variables at two time steps | FuXi-S2S Encoder-Perturbation-Decoder | / | 36 days |

**Table 1.** Summary of DL models forecasting the MJO. We only include studies that are totally based on data-driven methods and forecast RMM or snapshots that can be used to calculate RMM.

| Experiment | Input |
|---|---|
| BS | Eighteen variables (our **best** models) |
| CT | Single variable (**control** experiments for filtered-input experiments) |
| LA | Filtered single variable; **large-scale** signals in the meridional direction |
| LL | Filtered single variable; **large-scale** signals in both horizontal directions |
| SS | Filtered single variable; **small-scale** signals in both horizontal directions |

**Table 2.** Summary of five main sets of experiments. Two additional small-scale experiments are presented in the Supplementary Information (Fig. S20).

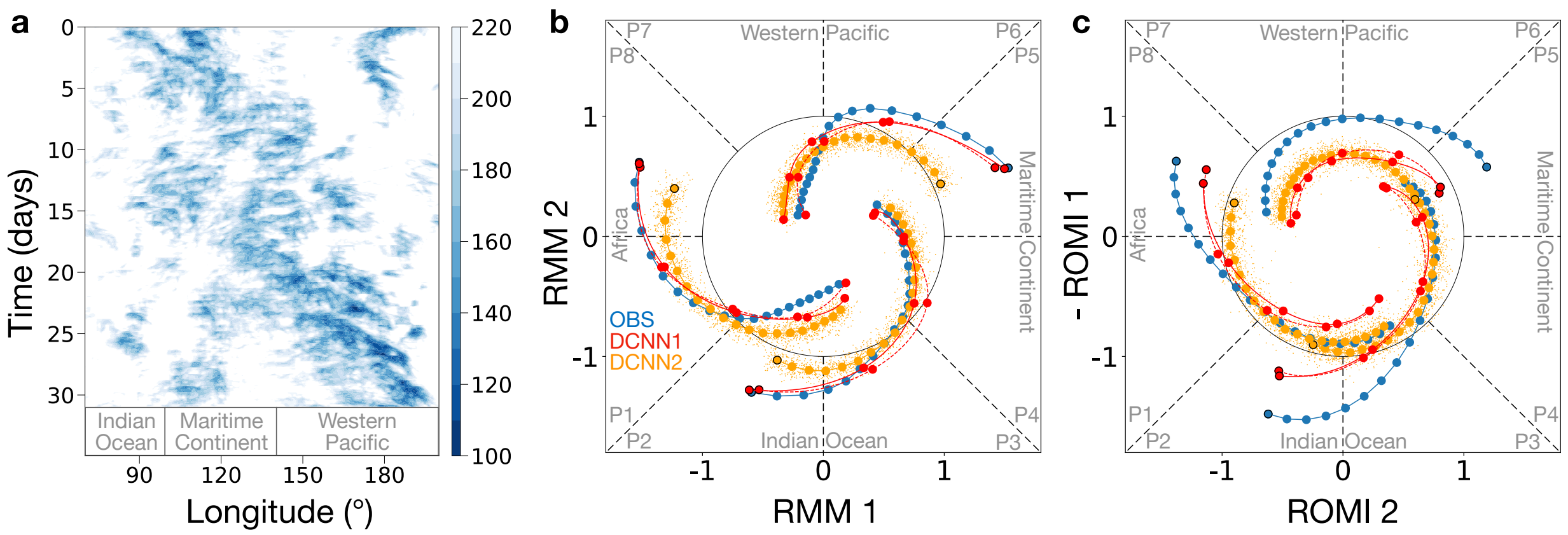

**Fig. 1. Observed and predicted evolution of the Madden-Julian oscillation (MJO).** (a) Longitude-time diagram showing an MJO event using outgoing longwave radiation (OLR) averaged between 0° - 10°S during December 1992. Blue shading indicates convective regions (OLR $\leq$ 220 W/m$^2$), while white indicates suppressed convection. High-resolution hourly ERA5 data (0.25° $\times$ 0.25°) reveal both the large-scale MJO convection slowly propagating eastwards and embedded small-scale convective systems moving in all directions. (b) Real-time Multivariate MJO index (RMM) diagram, illustrating the average MJO trajectories originating from different phases. Blue dots represent observed trajectories calculated from observational data (Methods). Predictions from two model configurations are shown: single-lead prediction models (DCNN1; red) and multi-lead prediction models (DCNN2; orange). Circles outlined in black mark initial conditions (day 0). Red circles represent five-day interval predictions interpolated using cubic spline curves, while orange circles represent daily predictions without interpolation. Solid and dashed red lines correspond to forecasts based on NOAA and ERA5 OLR data, respectively. Large circles represent ensemble means of deep learning (DL) forecasts, whereas smaller dots illustrate individual ensemble members (100 in total; Methods). MJO phases (P1–P8) and associated regions of enhanced convection are labeled in gray. Trajectories originating from other phases are shown in Fig. S1. (c) The Real-time OLR MJO index (ROMI) diagram, presented similarly to (b). Additional details on MJO indices and model setup are provided in Methods.

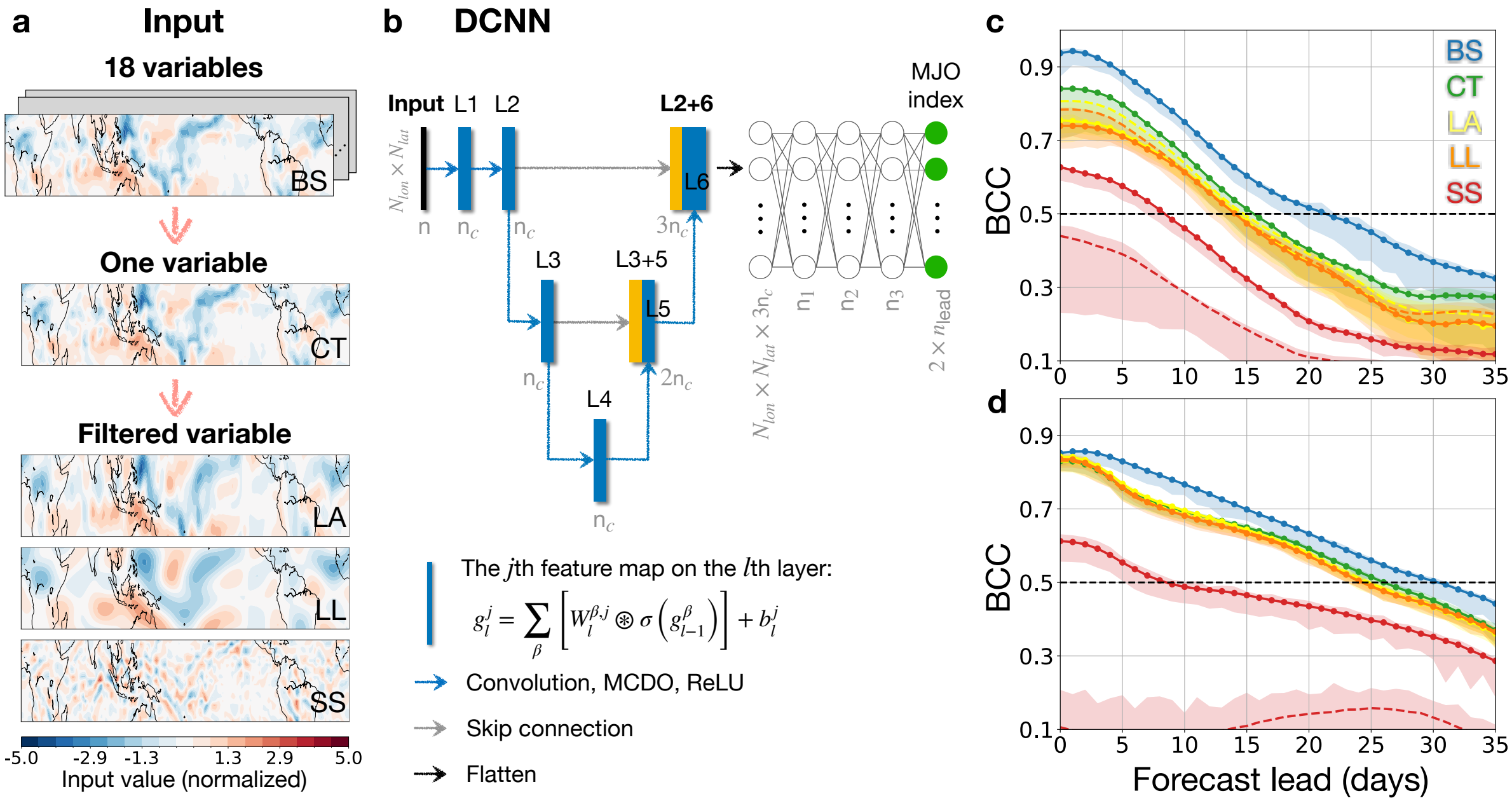

**Fig. 2. Deep convolutional neural network (DCNN) for MJO forecasting.** (a) Input examples from three levels of complexity used in five experiments, shown for NOAA OLR on 15 December 2021. Inputs are daily tropical snapshots (20°S–20°N, 2°×2° resolution): BS uses 18 variables, CT uses only one variable, and the rest use the spatially filtered variable. See Table 2 and Methods for details. (b) DCNN architecture overview. The black box is the input shown in (a). Blue boxes represent convolutional feature maps ($g_l^j$; Eq. (2)), with the number of feature maps per layer indicated in gray. Blue arrows indicate operations comprising convolution for feature extraction, Monte Carlo dropout (MCDO) for overfitting mitigation, and ReLU activation for nonlinearity. Grey arrows denote skip connections. Yellow boxes show copies of previous layers. Four fully connected layers (circles and lines) generate predictions of MJO indices using feature maps from layers L2 and L6 (Methods). (c) Performance of the five experiments for predicting the RMM using the multi-lead model DCNN2, evaluated by BCC. Solid lines indicate the ensemble-mean performance of models retrained from scratch. Dashed lines show the ensemble-mean performance when filtered inputs are directly applied to trained CT models without retraining. Shading indicates uncertainty estimated from 100 ensemble members (Methods). The dashed horizontal line denotes a BCC value of 0.5. (d) Similar to (c) but for predictions of ROMI.

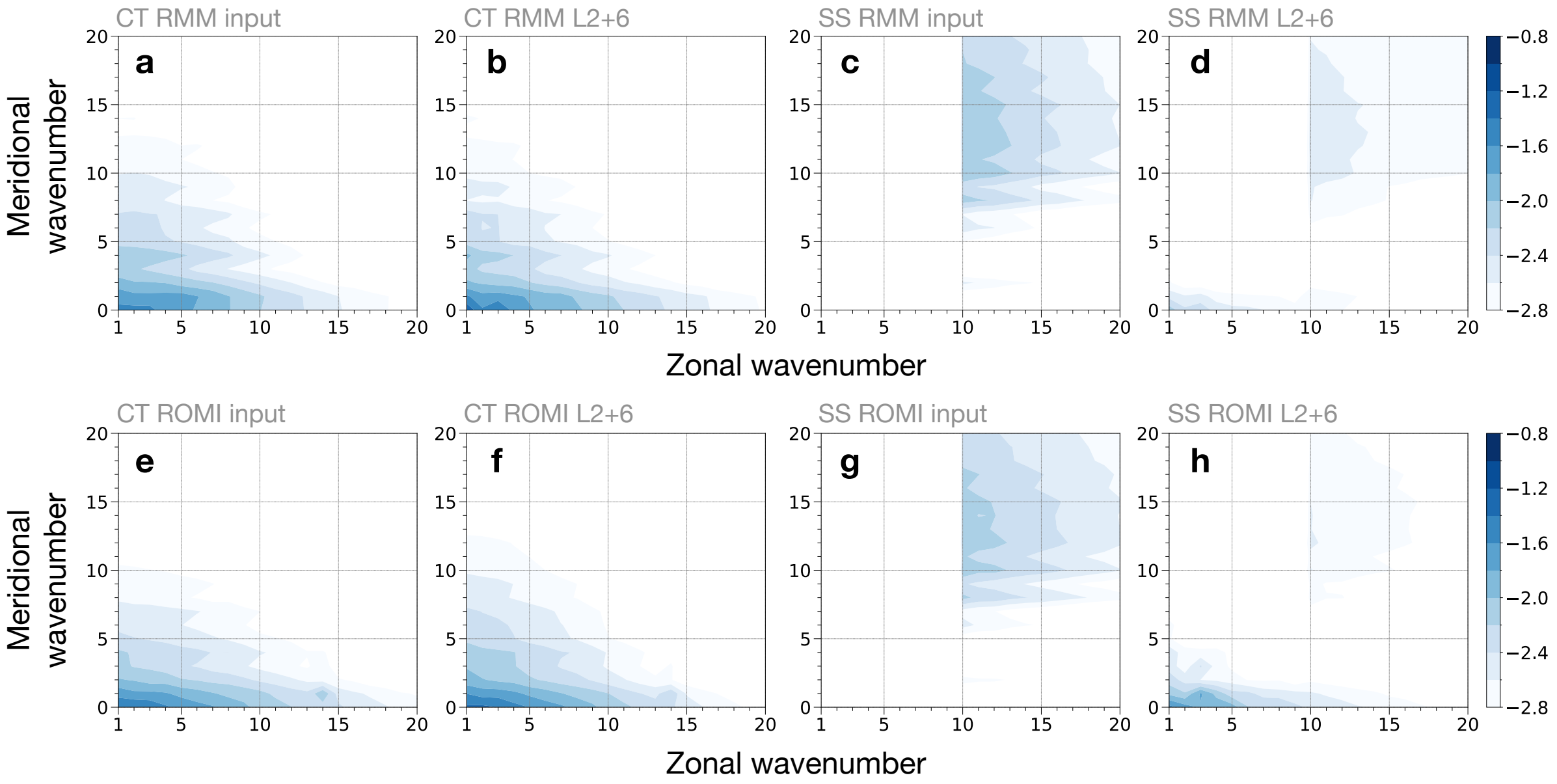

**Fig. 3. Spectral analysis of CT and SS models.** Panels show the log10-scaled, normalized power spectra of input fields and feature maps from the final convolutional layer (L2+6 before ReLU; $g^{j}_{2+6}$) in experiments CT and SS. The top row (a–d) corresponds to models predicting RMM, which use TCWV as input; the bottom row (e–h) corresponds to models predicting ROMI, which use OLR as input. Axes indicate zonal (x-axis) and meridional (y-axis) wavenumbers. Power is first normalized for each map by dividing raw spectral power by the total power across all wavenumbers. The spectra are then averaged over time, feature maps, and ensemble members. Colors represent normalized power on a logarithmic scale ($\log_{10}$), with values below –2.8 masked in white. Only a subset of feature maps, identified as most influential based on their individual contribution to the output, is selected for analysis. Details of the spectral analysis and selection procedure are provided in Methods.

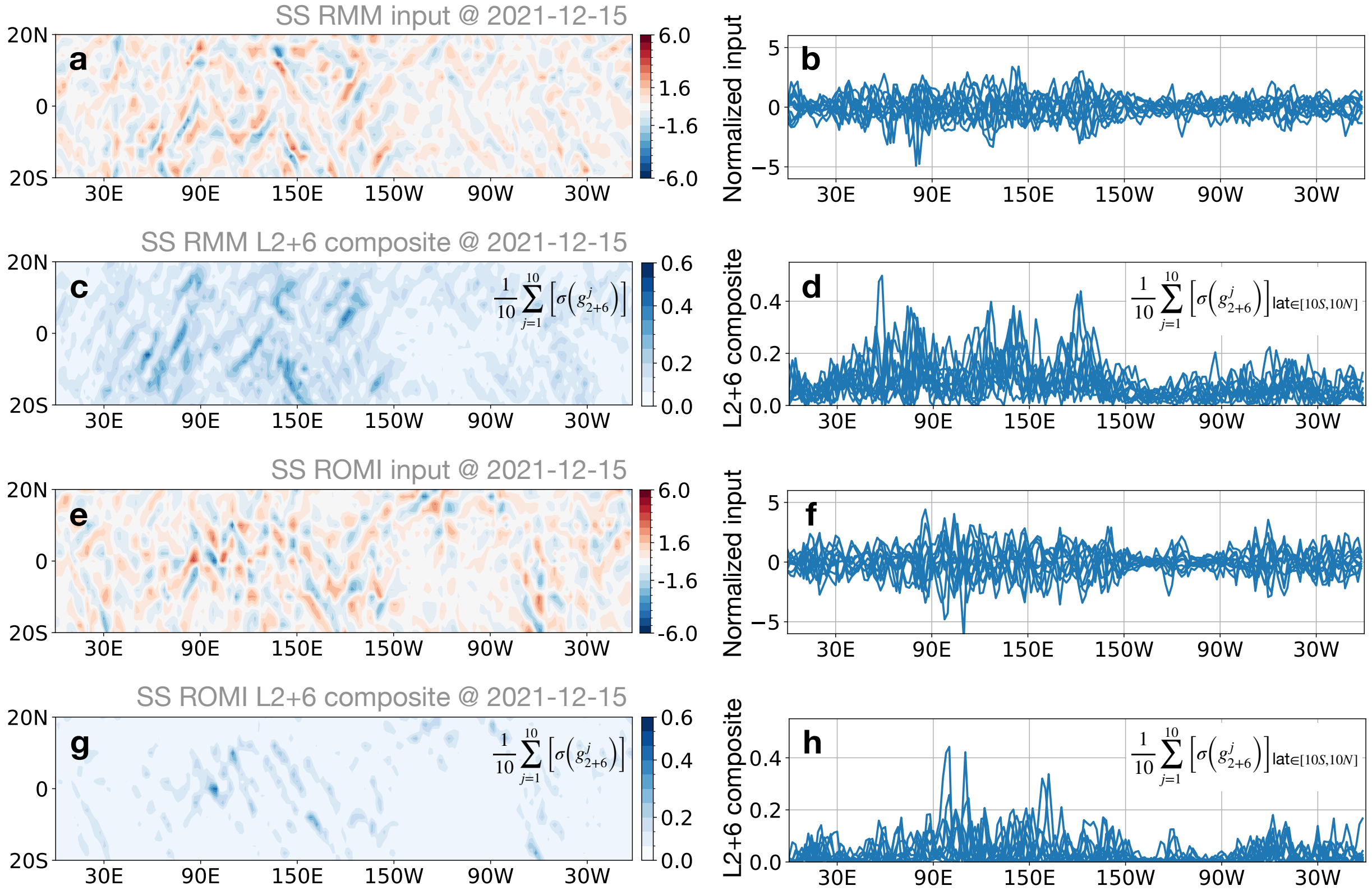

**Fig. 4. Input fields and composite feature maps in the small-scale (SS) experiment.** The top two rows correspond to models predicting RMM, and the bottom two rows to models predicting ROMI, all using data from 15 December 2021. (a, e) Input fields: TCWV for RMM and OLR for ROMI, each normalized by their global standard deviations. (b, f) Longitude profiles extracted from (a, e), taken along each latitude between 10°S and 10°N at 1° intervals. (c, g) Composite feature maps from the final convolutional layers (L2+6). The top 10 feature maps are selected based on their contribution to the output, determined by zeroing out individual channels (Methods). The composite is computed as the average across the selected feature maps after applying the ReLU activation: $\frac{1}{10}\sum_{j=1}^{10}\left[\sigma\left(g_{2+6}^{j}\right)\right]$. More details are provided in Methods and Fig. S26. (d, h) Longitude profiles extracted from (c, g), taken along each latitude between 10°S and 10°N at 1° intervals. Two distinct wave packets are visible in (d) and (h), separated at around 120°W and 30°E.

## Methods

**Data and preprocessing for the model input**. We use daily data from 1 January 1980 to 31 December 2021 to train (1980–2009), validate (2010–2015), and test (2016–2021) our DL models. OLR is taken from the NOAA interpolated OLR dataset at a 2° × 2° resolution[46], while all other atmospheric and oceanic variables are sourced from the ERA5 daily reanalysis dataset[47]. The final model input consists of 18 variables (BS), selected to represent the atmospheric and oceanic state. These include OLR, TCWV, specific humidity, zonal and meridional winds, air temperature, and geopotential height at three pressure levels (850, 500, and 200 hPa), along with sea surface temperature (SST).

All 18 variables are first averaged daily at four time points (00:00, 06:00, 12:00, and 18:00) with a spatial resolution of 2° × 2°. The daily data are then pre-processed to calculate anomalies by removing the time mean, the first three harmonics of the climatological seasonal cycle, and the previous 120-day averages at all grid points[31]. This preprocessing, following Wheeler and Hendon (2004), is designed to remove the seasonal cycle and the interannual variability, thereby highlighting the intraseasonal signals primarily associated with the MJO.

We also compared model performance using NOAA OLR versus ERA5 OLR and found no significant difference (Fig. 1b-c). Therefore, for simplicity and consistency, we use NOAA OLR for most experiments presented in this study without further notation. Results using ERA5 OLR are explicitly noted when applicable.

**MJO indices as the model output.** Two MJO indices, the real-time multivariate MJO index (RMM)[31] and the real-time OLR MJO index (ROMI)[32], are used as the model output to forecast the MJO. They are real-time indices based on empirical orthogonal function (EOF) analysis, an effective MJO filter without bandpass temporal filtering. Both indices have been used to evaluate the prediction skill of the MJO for models in the subseasonal-to-seasonal project[29-30].

RMM contains two values on each day, RMM1 and RMM2. These two values correspond to a point on the RMM diagram (Fig. 1b), where the phase indicates the geographic location of the MJO, and the distance to the origin indicates the magnitude of the MJO. RMM1 and RMM2 are the leading pair of principal components (PCs) of the combined fields of meridionally averaged anomalous OLR, 850-hPa zonal wind, and 200-hPa zonal wind in the tropics (15°S–15°N). These daily anomalous fields are calculated following Wheeler and Hendon (2004), similar to the preprocessing process for the models' input. Each field is normalized by its global variance. We calculate the EOFs of the combined anomalous fields using data from 1979 to 2001. Then, we project the combined fields of the whole period onto the leading pair of multiple-variable EOFs to get the PC time series. RMM1 and RMM2 are normalized by their standard deviations from 1979 to 2001.

Similarly, ROMI uses two values, ROMI1 and ROMI2, to indicate the location and magnitude of the MJO (Fig. 1c). Unlike RMM, ROMI is the leading pair of PCs of the anomalous OLR field, characterized by the rainstorms of the MJO. OLR is much noisier than zonal winds and requires more filtering. Following Kiladis et al. (2014), we first remove the time mean and the first three harmonics of the climatological seasonal cycle from the daily OLR to calculate OLR anomalies. Next, we subtract the previous 40-day averages from the OLR anomalies to remove low-frequency variability and apply 9-day tapered running averages to remove intraseasonal variability. Then, we project these processed OLR anomalies onto the EOF patterns, which are calculated for each day of the year using the 30-to-96-day eastward-filtered OLR from 1979 to 2012, employing the Python package developed by Hoffmann et al. (2021). ROMI1 and ROMI2 are both normalized by the standard deviation of ROMI1. When visualized in the ROMI phase space, ROMI2 corresponds to RMM1, and –ROMI1 corresponds to RMM2, enabling direct comparison between the two indices.

**Model architectures.** Three model architectures are explored in this study (Figs. 2b, S9, and S32a). We start with a DCNN architecture with a U-Net-like shape (Fig. 2b). U-Net was first used for medical image segmentation[49] and has been successfully applied to many other fields, such as weather forecasts for large-scale atmospheric circulations[50]. In addition to an encoding path performing convolutions typical to CNN, the U-Net contains a decoding path that concatenates copies from the previous shallow layers to the deeper layers after convolution (Grey arrows in Fig. 2b).

In our DCNN setup (Fig. 2), the input channels are daily snapshots of variables in the tropics (20°S–20°N), each channel with a size of 180 (longitude points $N_{lon}$) × 21 (latitude points $N_{lat}$). The input channels can be eighteen or one, depending on the experiment design (Fig. 2a and Table 2). The output is given by RMM or ROMI, which are two scalar quantities for individual leads. We implement two model configurations: (1) single-lead models (DCNN1), which predict the MJO index at a single lead time; (2) multi-lead models (DCNN2), which predict the index at multiple lead times simultaneously. The only difference lies in their loss functions.
The loss function for single-lead models is:

$$L_{single}(\tau) = \frac{1}{N}\sum_{i=1}^{N} \|P_{i,\tau} - T_{i,\tau}\|^2,$$

where $\tau$ is the lead time, $N$ is the batch size, $P_{i,\tau}$ is the $ith$ predicted MJO index at lead $\tau$ in the batch. $T$ is the target. Similarly, the loss function for multi-lead models is:

$$L_{multi} = \frac{1}{NL}\sum_{i=1}^{N}\sum_{\tau=0}^{L} \|P_{i,\tau} - T_{i,\tau}\|^2.$$

$L$ is set to 35 days in the main text. The model performance is not sensitive to $L$ (Fig. S34).

Between the input and output, there are four convolutional layers in the encoding path and two convolutional layers in the decoding path, followed by four fully connected layers. In the encoding pathway (from input to L4), channels on each convolutional layer are convolved with filters at a stride of 1, with zero padding at the edges to generate feature maps of the same size as the input channels. The $j$th feature map on the $l$th convolutional layer at a given time step is given by:

$$g_l^j = \sum_{\beta}\left[W_l^{\beta,j} \circledast \sigma\left(g_{l-1}^{\beta}\right)\right] + b_l^j, \qquad g_l^j \in \mathbb{R}^{N_{lon}\times N_{lat}}, \tag{2}$$

where $W_l^{\beta,j}$ is the kernel weight matrix, $\beta$ indexes the feature maps of the previous layer, $\circledast$ is the convolution, $\sigma(\cdot) = \max(0,\cdot)$ is the ReLU activation function, and $b_l^j$ is the bias. These feature maps are passed through the Monte Carlo Dropout (MCDO) layer, where 10 percent of the channels are randomly selected and zeroed out during the training stage. The output of the MCDO layer is then scaled with $\frac{1}{1-0.1}$ to maintain the magnitude of the total output. MCDO is disabled during evaluation. MCDO is designed to prevent overfitting and improve the model's performance. Each convolutional layer is also followed by a ReLU activation function that includes nonlinearity. There is NO maxpooling, upsampling, and batch normalization in our DCNN.

In the decoding path, the size of each feature map is the same, but the number of feature maps increases by concatenating copies from shallower layers (e.g., L3 and L2) to the deeper layers (e.g., L5 and L6) after convolution (Fig. 2b). These concatenated feature maps are similarly passed through the MCDO layer and the ReLU activation function. Then, the feature maps on the last convolutional layer (L2+6) are flattened into a long vector and passed to four fully connected layers. The fully connected layers progressively reduce the feature map dimensions, producing the predicted RMM or ROMI at a given or multiple lead times. Each fully connected layer is followed by an MCDO layer and a ReLU activation function. The loss function is the mean squared error (MSE), which measures the average squared differences between the predicted and actual MJO indices.

Fig. S9 shows the second architecture: a complex DCNN variant, which uses a standard U-Net architecture to predict RMMs at 8 lead times in one single model. Details on the model setup are described in its caption.

Fig. S32a shows the third architecture: a two-layer CNN. Its model setups are the same as DCNN, except that we remove the convolutional layers after L2. This reduction lowers the number of trainable parameters in CT models from 436 million to 145 million for RMM prediction, and from 73 million to 24 million for ROMI prediction.

**Metrics to quantify the model performance**. We employ the following two metrics to evaluate our DL models' performance. The bivariate correlation coefficient (BCC) assesses the correlation between the predicted and observed MJO indices. It provides a general measure of the strength and direction of the linear relationship between the two variables. A higher BCC indicates a more robust and accurate prediction of the MJO indices. The root mean square error (RMSE) quantifies the average magnitude error between the predicted and observed MJO indices. Lower RMSE values indicate better predictive performance. These metrics provide a general evaluation of the model's accuracy and reliability in forecasting the MJO. They have been extensively used in previous research[25-30], helping us compare the performance of different models.

**Model selection.** To optimize model performance, we conduct hyperparameter tuning using the Optuna framework. The search space includes learning rate (log-uniform from $10^{-4}$ to $10^{-2}$), batch size (10 to 100, in steps of 10), dropout rate (0.1 to 0.5), kernel size (3, 5, or 7), and optimizer choice (SGD or Adam). We also tune the number of filters in the encoder (16 to 64, in steps of 16), as well as the number of neurons in the three fully connected layers (ranging from 100 to 600, 50 to 250, and 10 to 100, respectively). Model performance is evaluated using the BCC on the validation set, and the model with the highest validation BCC is selected. We implement early stopping with a patience of 10 epochs to prevent overfitting. The best hyperparameter configurations are summarized in Table S1. All code used in this study is publicly available at https://github.com/linyao1999/DL4MJO_UNet.git.

**Spectral analysis.** For this study, we take $c = 51$ m s$^{-1}$ and $\beta = 2.28 \times 10^{-11}$ m$^{-1}$s$^{-1}$ in Eq. (1), yielding an equatorial deformation radius of 1496 km. Changing $c$ to 20 m s$^{-1}$ produces consistent results (Fig. S19). Each power spectrum $\hat{q}(k,m)$ is normalized by its total power $\int_0^{+\infty}\int_{-\infty}^{+\infty} \hat{q}^2(k,m)dkdm$. The normalized spectra are then averaged across feature maps and time steps. We exclude the zonal wavenumber $k = 0$ from the calculation, as the MJO is a propagating signal. The average normalized power spectra for the CT and SS models are shown on a log scale in Fig. 3. We also compare how these power spectra evolve from the input to the last convolutional layer in the Supplementary Information (Figs. S17 and S25) to further support the conclusion drawn from Fig. 3. Note that the meridional projection is imperfect because $y$ cannot reach to infinity in Eq. (1). We have tested the sensitivity of our experiments to the imperfect meridional projection by using different latitudinal ranges for input (Supplementary Text 1 and Fig. S18). The results confirm that the findings from our experiments are consistent regardless of the latitude range used, thus robust to imperfections in the meridional projection.

**Experimental setup**. To investigate the spatial scales most relevant to MJO predictability, we design a hierarchy of DL experiments using multi-lead models (DCNN2), summarized in Fig. 2a and Table 2. The first experiment (BS) uses the full set of 18 daily tropical fields as input, representing a comprehensive description of the atmospheric and oceanic state. The second set of experiments simplifies the input to a single variable. Among all candidate variables, we find that OLR yields the highest prediction skill for ROMI, while TCWV performs best for RMM. These one-variable models form the control experiment (CT) for the subsequent filtered-input experiments.

In the third set of experiments (LA, LL, and SS), we further simplify the CT input by applying spatial filtering to isolate or remove specific spatial scales. The large-scale experiments remove small-scale information: LA filters in the meridional direction only, while LL applies filtering in both meridional and zonal directions. The small-scale experiment (SS) retains only small-scale signals by removing large-scale components in both directions. To further investigate scale interactions, we include two additional small-scale experiments in the Supplementary Information (SL and SA; Fig. S20). SL retains large-scale meridional structures while applying small-scale zonal filtering, and SA applies small-scale zonal filtering without meridional filtering. In all filtered experiments, the cutoff wavenumber is set to 10 in both directions, corresponding to a zonal wavelength of ~4,000 km, based on spectral analysis of input and feature maps (Fig. 3). To quantify uncertainty, each experiment is repeated with an ensemble of 100 independently trained models, each initialized with different random weights.

In addition to the multi-lead configuration using NOAA OLR and ERA5 variables, we conduct complementary experiments using single-lead models to validate key findings. For example, experiment BS is run with both NOAA and ERA5 OLR, using 100 ensemble members at eight lead times (Figs. 1b–c and S1–S6). Experiment CT is also run with ERA5 OLR to identify which variable gives the best forecasts, using 100 ensemble members (Fig. S16). Filtered-input experiments (LA–SA) are repeated with ERA5 OLR using 11 ensemble members (Figs. S24, S29–S31, S33). While the ensemble size is reduced for computational efficiency, the results remain robust. Overall, the single-lead models produce results consistent with those of the multi-lead models, with the added benefit of improved performance at shorter lead times.

**Evaluation of the contribution from each feature map of the last convolutional layer**. To evaluate the contribution of each feature map of the last convolutional layer, we zero out each feature map each time and calculate the average difference between the new output and the original output. Then, we rank the feature maps based on the average difference of the output. The result is shown in Figs. S28-29. This method follows Shin et al. (2022). The difference is that we zero out the entire feature map instead of a square region in the feature map each time.

We also perform the sensitivity analysis to further confirm the contribution evaluation. Instead of zeroing out each feature map, we add small perturbations at each grid point for each feature map each time and calculate the changes in the output. The small perturbation is defined as the product of one standard deviation of the feature map at each grid point and a weight between 0 and 1 given by a Gaussian distribution. Similarly, we calculate the average difference between the new and original output. Due to the computational cost of this method, we apply it only to the single-lead models using ERA5 OLR as input. The result is shown in Supplementary Information (Fig. S30). This sensitivity analysis shows how sensitive the output is to the small changes in the hidden feature maps.

**Feature map composite**. To visualize the most influential feature maps, we construct a composite of feature maps from the last convoluti onal layer by selecting the top 10 feature maps that contribute most to the output[36] (Fig. S26) The composite is computed by taking the mean value of the selected feature maps at each grid point, defined as:

$$\frac{1}{10}\sum_{j=1}^{10}\left[\sigma\left(g_{2+6}^{j}\right)\right],$$

This approach highlights the key spatial patterns captured by the most important feature maps. Results are shown in Figs. 4 and S27.

**Data availability**

The ERA5 hourly data are available from https://cds.climate.copernicus.eu/datasets/reanalysis-era5-pressure-levels and https://cds.climate.copernicus.eu/datasets/reanalysis-era5-single-levels. The NOAA Interpolated Outgoing Longwave Radiation (OLR) reanalysis data were downloaded from https://psl.noaa.gov/data/gridded/data.olrcdr.interp.html.

**Code availability**

The codes used to produce the results are available at https://github.com/linyao1999/DL4MJO_UNet.git.

The Python package[45] to calculate the OMI EOFs is available at https://github.com/cghoffmann/mjoindices.

**Acknowledgements**
This research was supported by a Packard Fellowship awarded to D.Y. and NSF grant AGS-2531264 to P.H. We gratefully acknowledge the computing resources provided by NERSC. We also thank Zhihong Tan and colleagues at the University of Chicago (e.g., Tiffany Shaw, Noboru Nakamura, Y. Qiang Sun, Yaoxuan Zeng, and Giorgio Sarro) for valuable feedback on the research.

**Author contributions**
D.Y. conceptualized and supervised the research. L.Y. conducted the majority of the experiments presented in the study, while J.D. carried out the remaining experiments. J.D. and A.C. assisted with model training. L.Y. and D.Y. drafted the initial version of the manuscript. All other co-authors contributed by offering feedback, suggesting experimental designs, and collaboratively revising the manuscript through multiple rounds.

**Corresponding author**
Correspondence to Lin Yao and Da Yang.

**Competing interests**
The authors declare no competing interests.

**Supplementary Information for**
**"Deep learning the sources of MJO predictability: a spectral view of learned features"**

This file includes:

Texts S1 to S2
Figures S1 to S33
Table S1

**Supplementary Text 1. Sensitivity of filtered-input experiments to meridional projection imperfections**

We assess the robustness of our filtered-input experiments to potential imperfections in the meridional projection by expanding the input domain from 20°S–20°N to 40°S–40°N. As shown in Figure S18, the results remain consistent with those presented in Figures 2c–d. This broader input domain marginally improves the performance of the large-scale experiments (LA and LL) while slightly reducing the performance of the small-scale experiment (SS). These changes are minor and do not alter the key conclusions drawn from the original experiments. The results confirm that our findings are robust to variations in the meridional projection domain. We also conduct similar experiments using single-lead models with ERA5 OLR, which yield consistent results (not shown).

**Supplementary Text 2. Sensitivity of model performance to cutoff wavenumbers**

To assess how the choice of spatial cutoff affects model performance, we vary the zonal and meridional cutoff wavenumbers applied to the input fields in the filtered-input experiments (LA–SS; Figures S22–S23). All models are retrained from scratch using the corresponding filtered inputs. In the multi-lead setup, performance in the large-scale experiments (LA and LL) improves modestly with higher cutoff wavenumbers, while performance in the small-scale experiment (SS) declines more steeply. Across all cutoffs, large-scale experiments consistently outperform small-scale ones.

To better capture prediction skill at shorter lead times—where multi-lead models may be less optimal—we also conduct experiments using single-lead models (Figure S24). These models show similar trends: large-scale performance increases with relaxed cutoffs, while small-scale performance declines. Notably, when the zonal cutoff is reduced to 5, small-scale models begin to outperform their large-scale counterparts. These results suggest that the relative importance of spatial scales is sensitive to the definition of “large” and “small”. When we define cutoffs based on the MJO theories (e.g., m=3, k=3), small-scale signals can be more effective than large-scale signals.

**Figures**

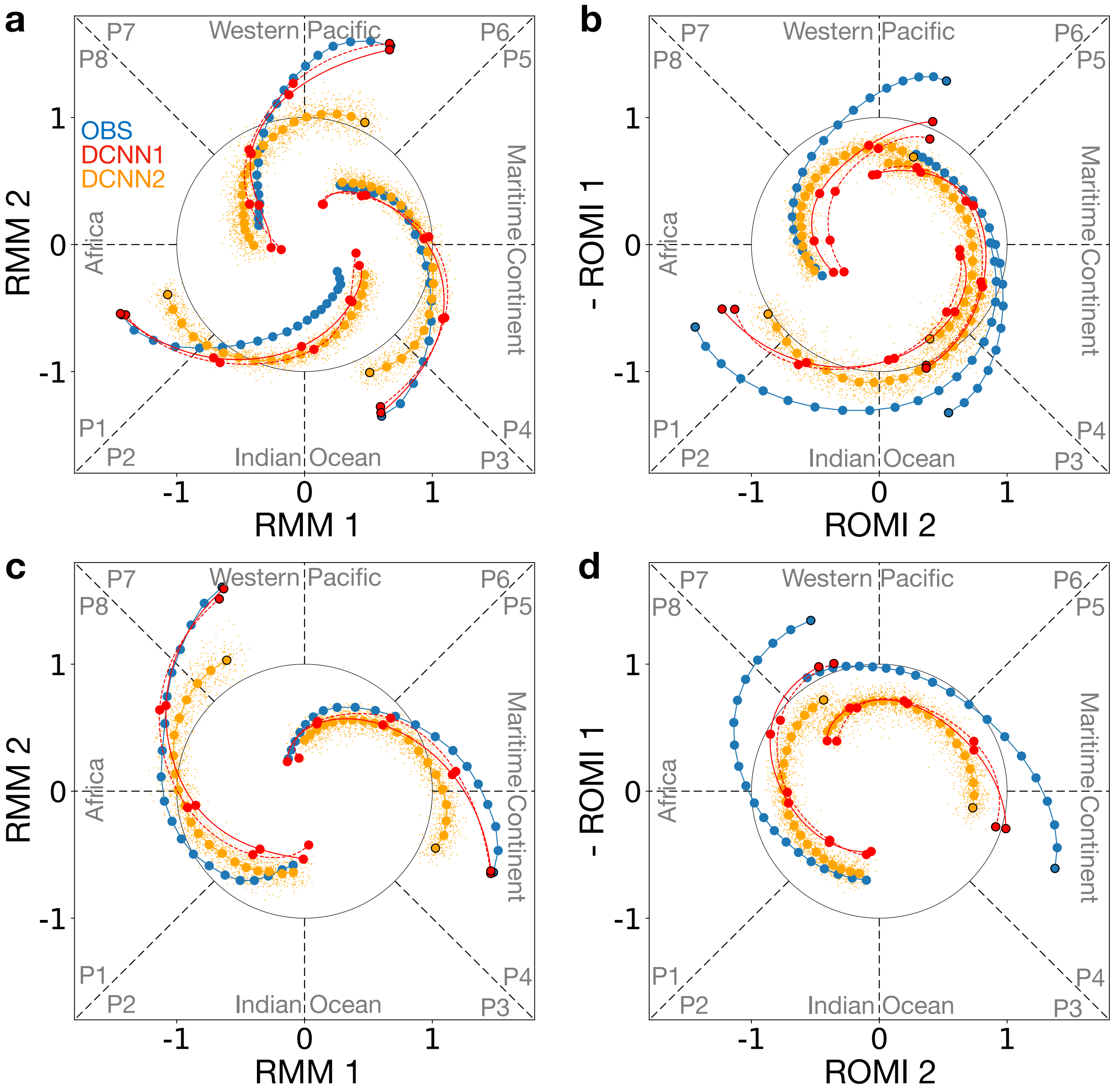

**Figure S1.** Same as Figures 1b and 1c, but for MJO trajectories originating from the other phases.

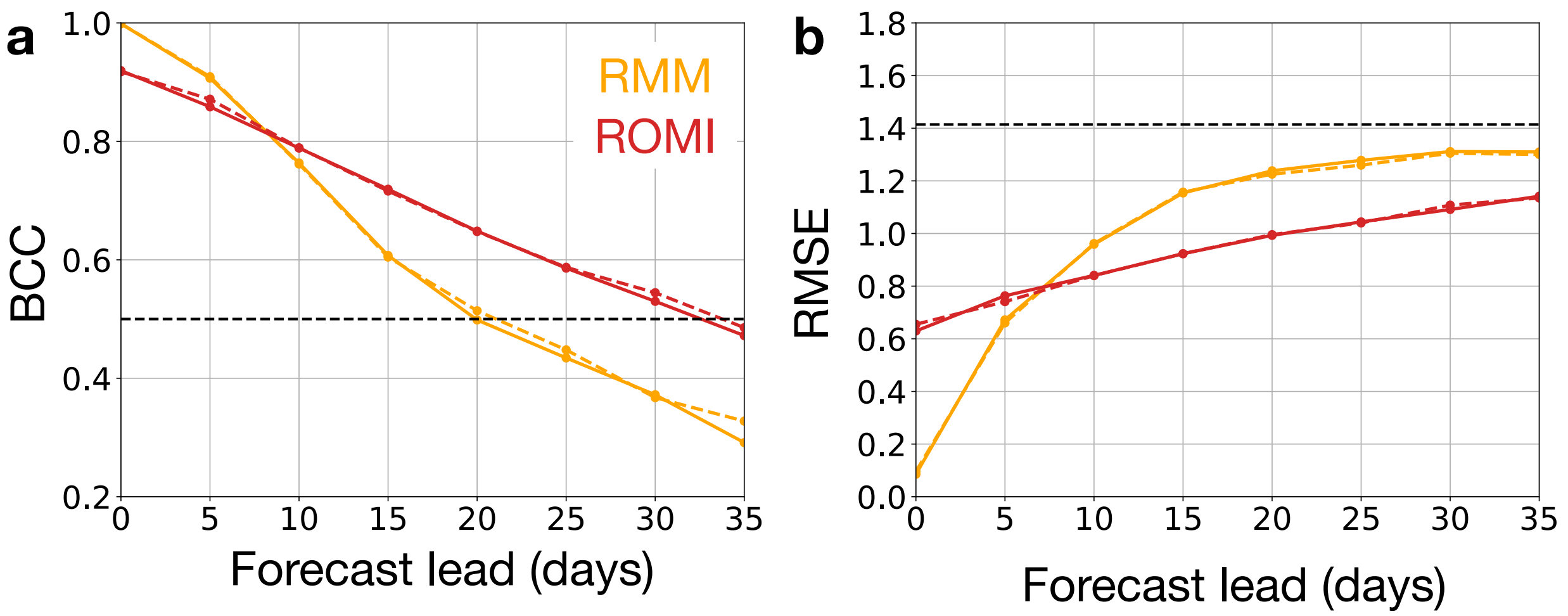

**Figure S2.** Performance of experiments BS in the single-lead model DCNN1. Solid lines use NOAA OLR as the model input and to compute the MJO index, while dashed lines use ERA5 OLR. (a) BCC. (b) RMSE.

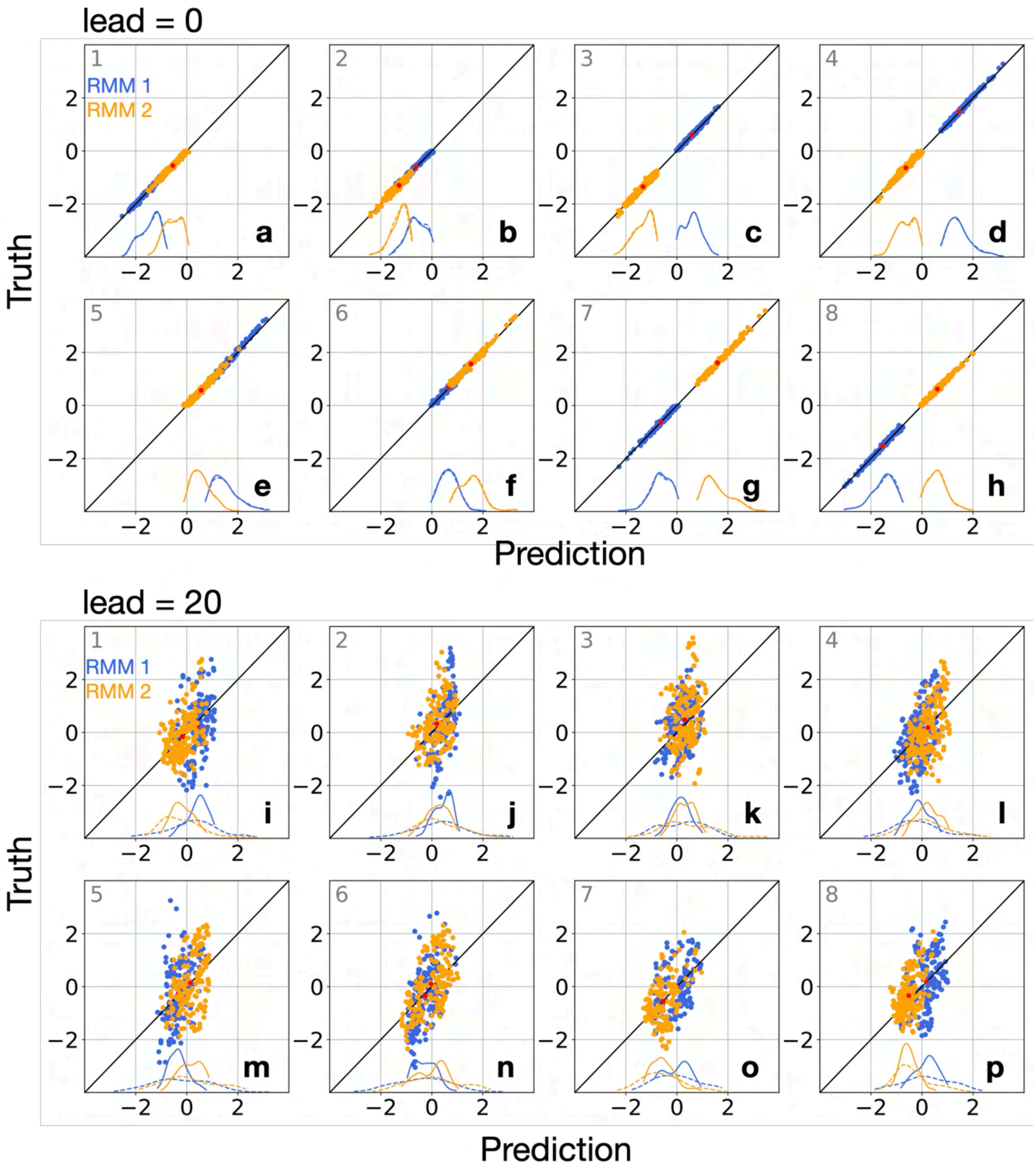

**Figure S3.** One-to-one comparisons between predicted and observed RMM indices from the single-lead models (DCNN1) using NOAA OLR and 17 additional ERA5 variables as input. Panels (a–h) show predictions at lead day 0, with each panel corresponding to MJO events initialized in one of the eight phases (indicated in the upper left of each panel). The x-axis shows the predicted RMM values, and the y-axis shows the observed values. Blue and orange dots represent RMM1 and RMM2, respectively, with red dots indicating the cluster centers. The black diagonal lines denote the one-to-one reference. The closer the dots are to the line, the better the prediction is. We also show the distribution of RMM1 and RMM2 at the bottom of each panel,

where the solid lines represent the predicted RMM1 (blue) and RMM2 (orange), and the dashed lines represent the true RMM1 (blue) and RMM2 (orange). At this lead, the predicted RMMs accurately capture the distribution of the true RMMs. Panels (i–p) show the same diagnostics for lead day 20. The predicted RMMs are more concentrated than the true RMMs. Although the mean predicted RMMs are close to the mean true RMMs (red dots are close to the one-to-one line), the DCNN1 model fails to accurately capture the tails of the true RMM distribution, which has a larger variation.

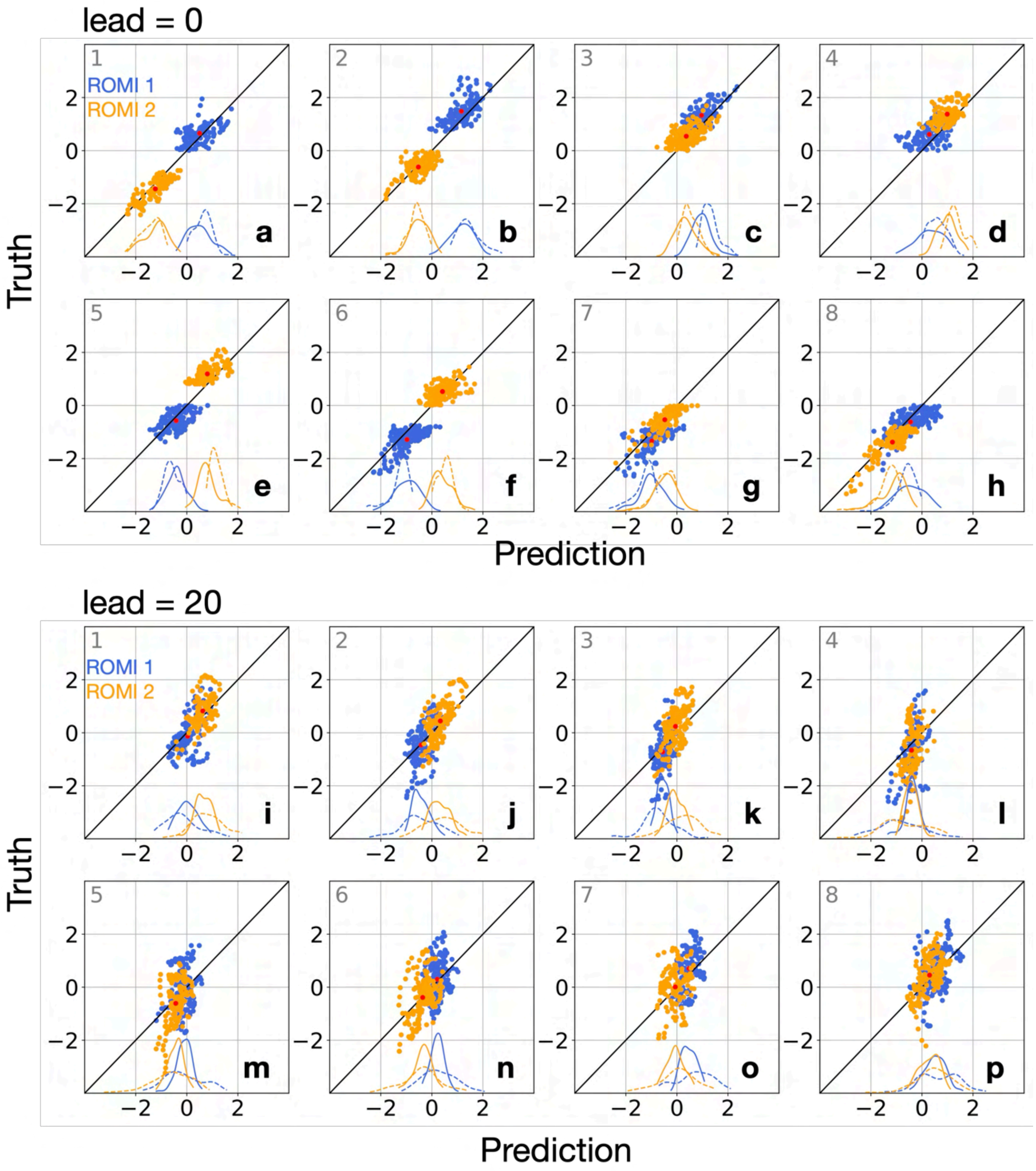

**Figure S4.** Same as Figure S3, but for models predicting ROMI.

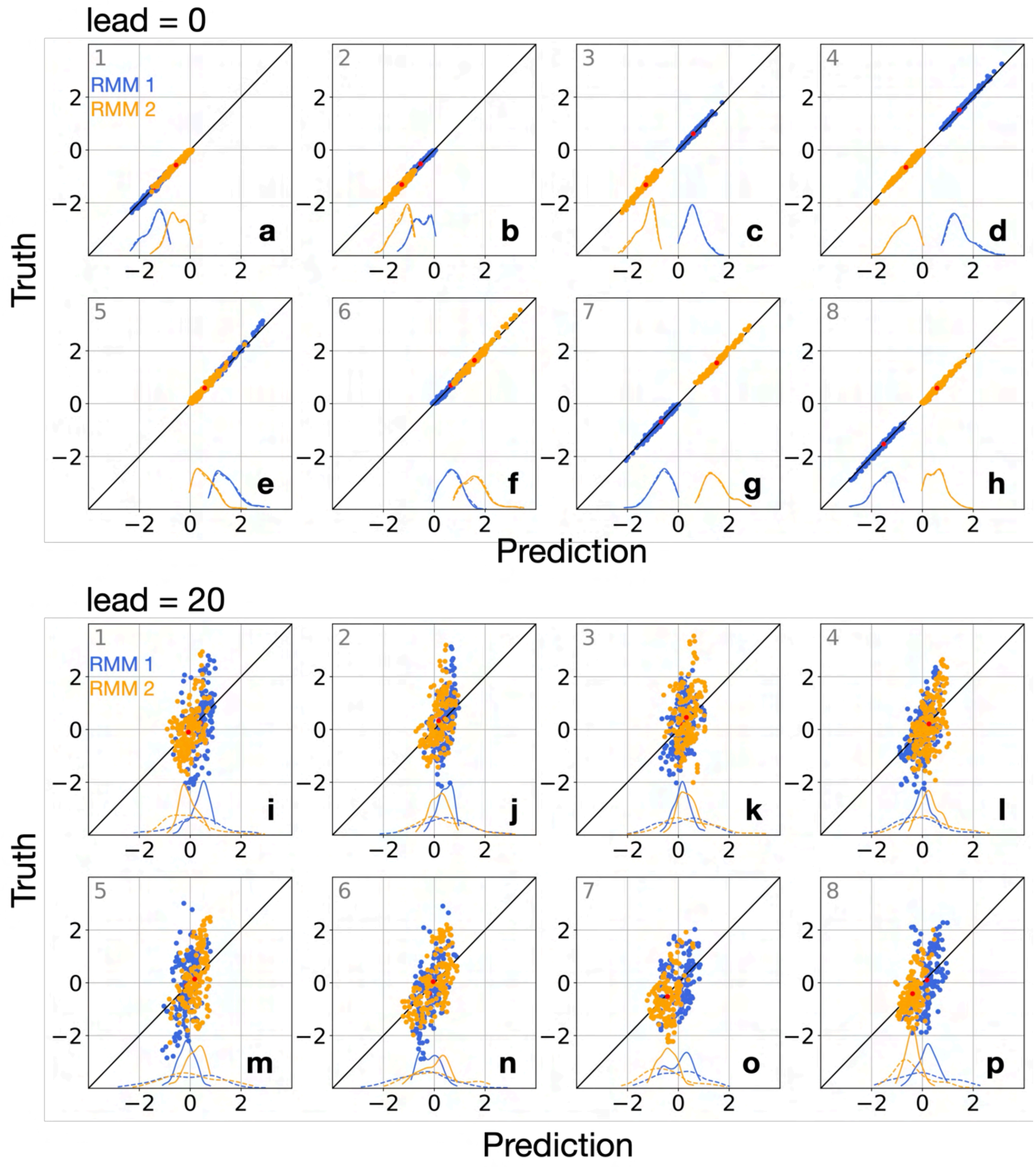

**Figure S5.** Same as Figure S3, but for single-lead models (DCNN1) using OLR and the remaining 17 variables from ERA5 as input.

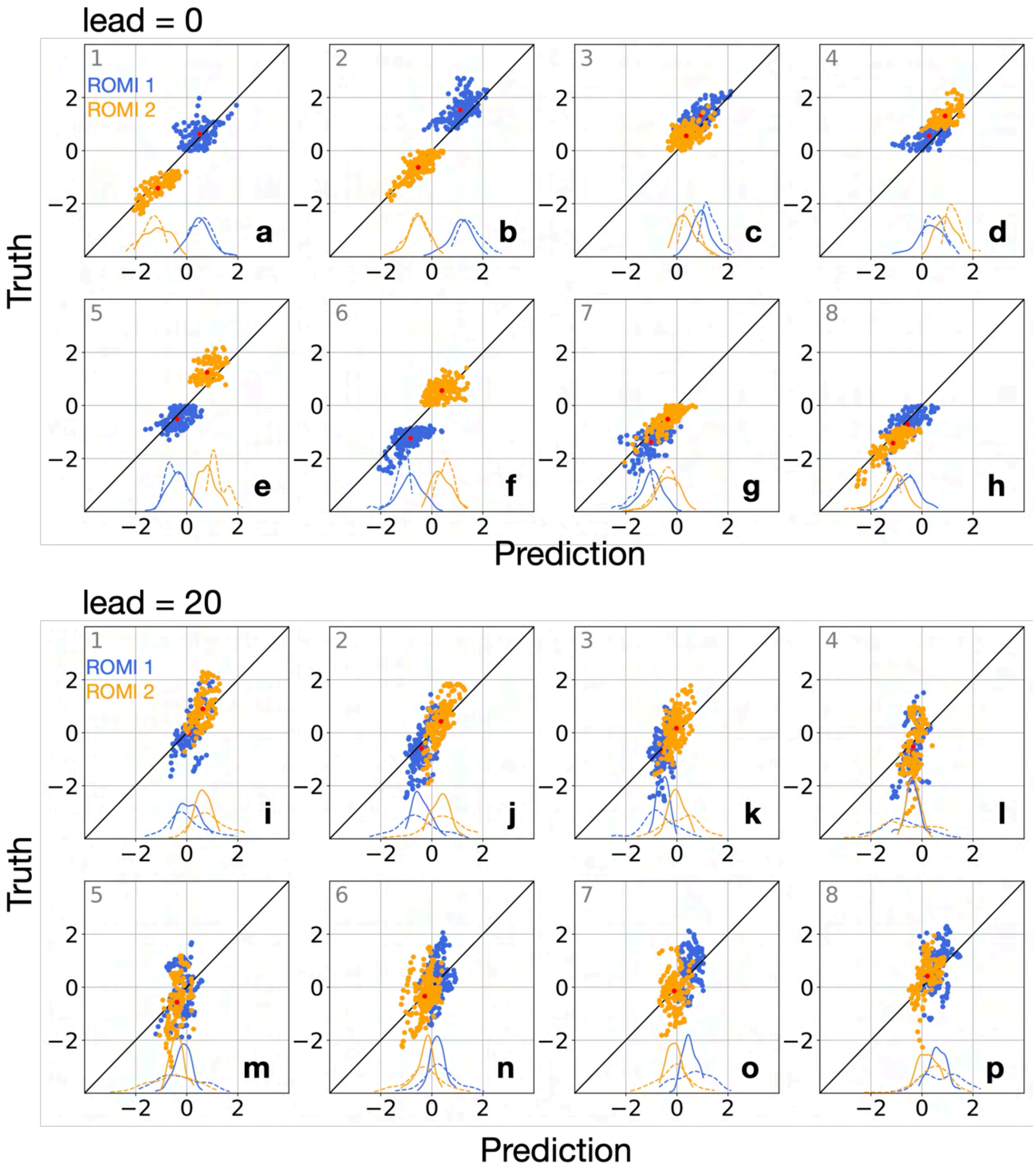

**Figure S6.** Same as Figure S5, but for models predicting ROMI.

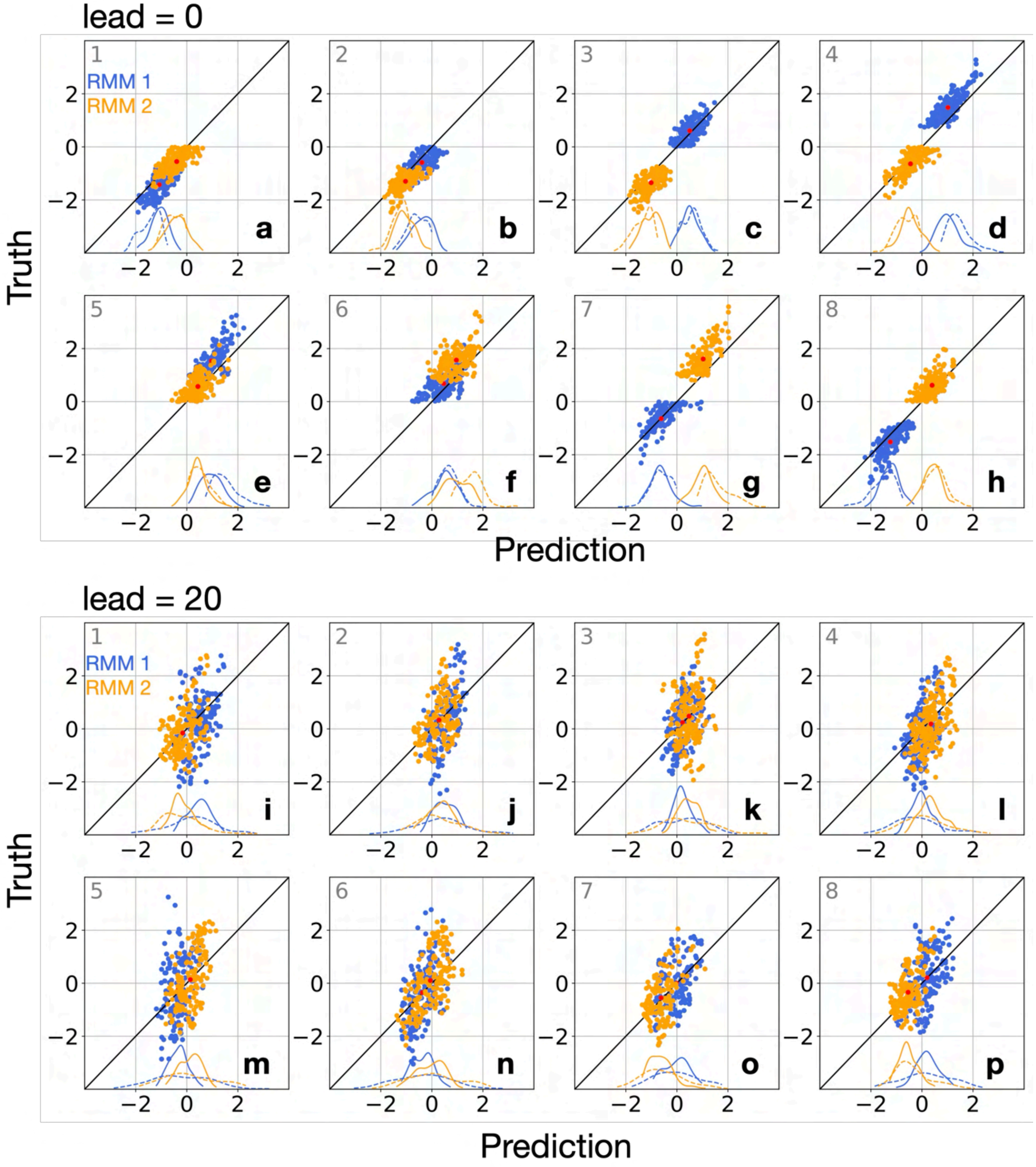

**Figure S7.** Same as Figure S3, but for multi-lead models (DCNN2) predicting RMM.

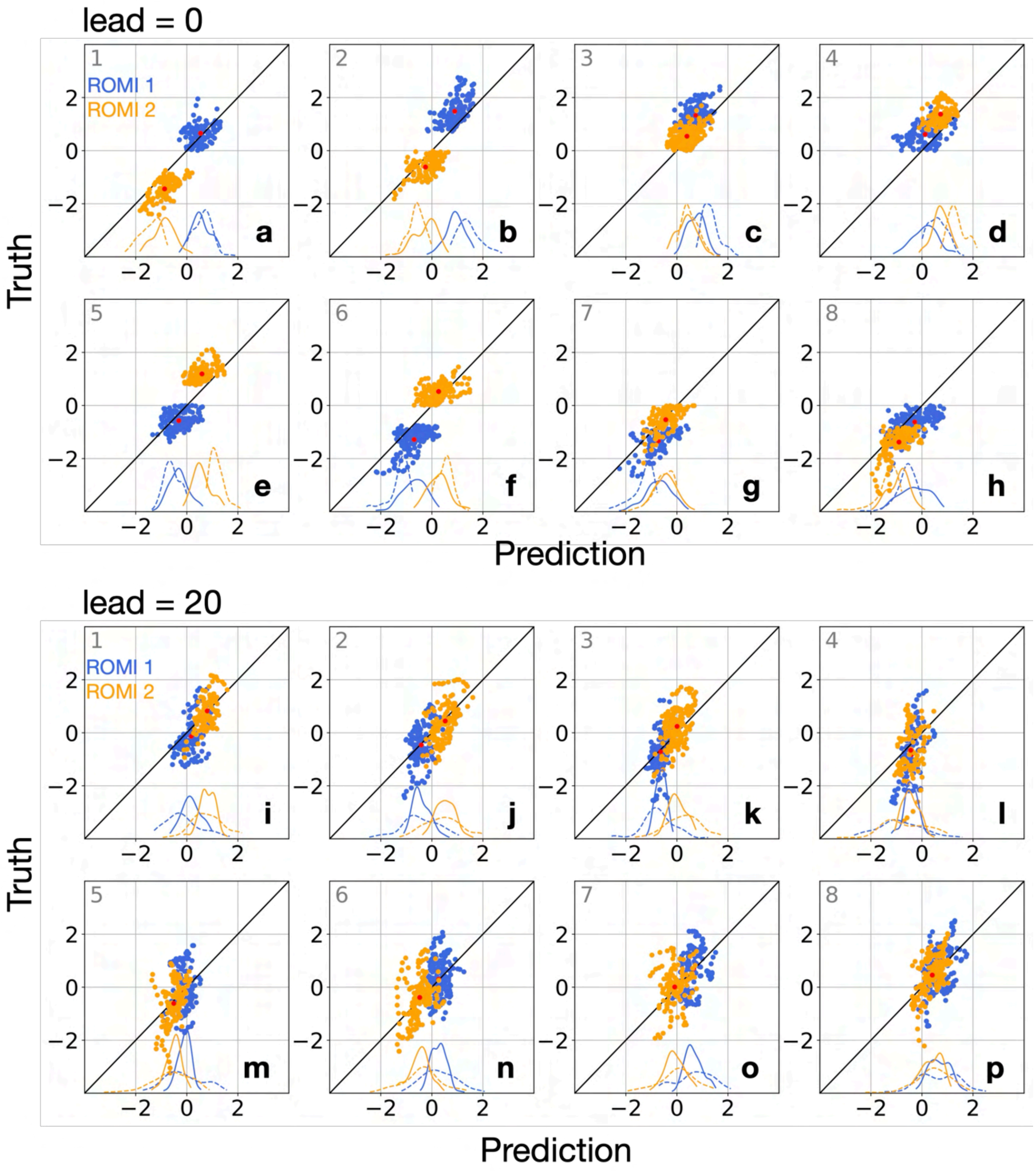

**Figure S8.** Same as Figure S7, but for models predicting ROMI.

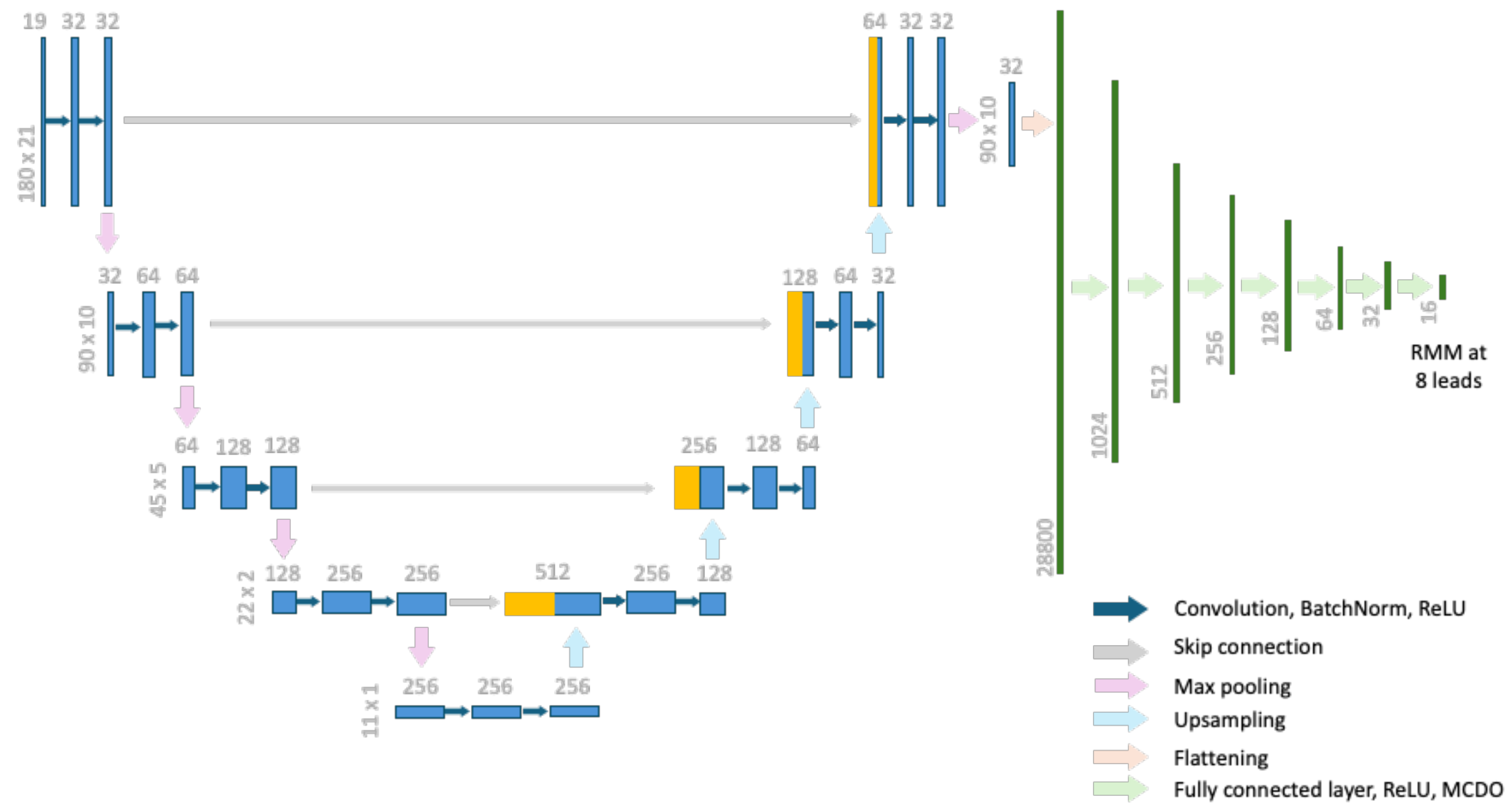

**Figure S9.** Architecture of the standard U-Net used to predict RMM at 8 lead times simultaneously. The left panel depicts the U-Net architecture, while the right panel shows the fully connected layers (FCNs). The model takes 19 input fields of size 180 × 21 and processes them through a series of convolutional, downsampling, and upsampling operations to produce RMM predictions. Each convolutional layer comprises a 3×3 convolution followed by batch normalization and a ReLU activation (dark blue arrows), with two such layers forming each convolutional block. Max pooling layers (pink arrows) reduce spatial resolution by a factor of two. Upsampling layers (light blue arrows) apply bilinear interpolation to double spatial resolution, followed by additional convolutional blocks. Skip connections (gray arrows) transfer spatial information from the encoder to the decoder via feature map concatenation (orange boxes). The output is flattened (peach arrows) and passed through FCNs (green arrows) with ReLU activations and Monte Carlo Dropout (MCDO) before producing the final prediction. The model is trained using a cosine annealing learning rate schedule that decays from $1.5\times10^{-4}$ to zero over 60 epochs. The batch size is 64. Numbers above each box indicate the number of channels.

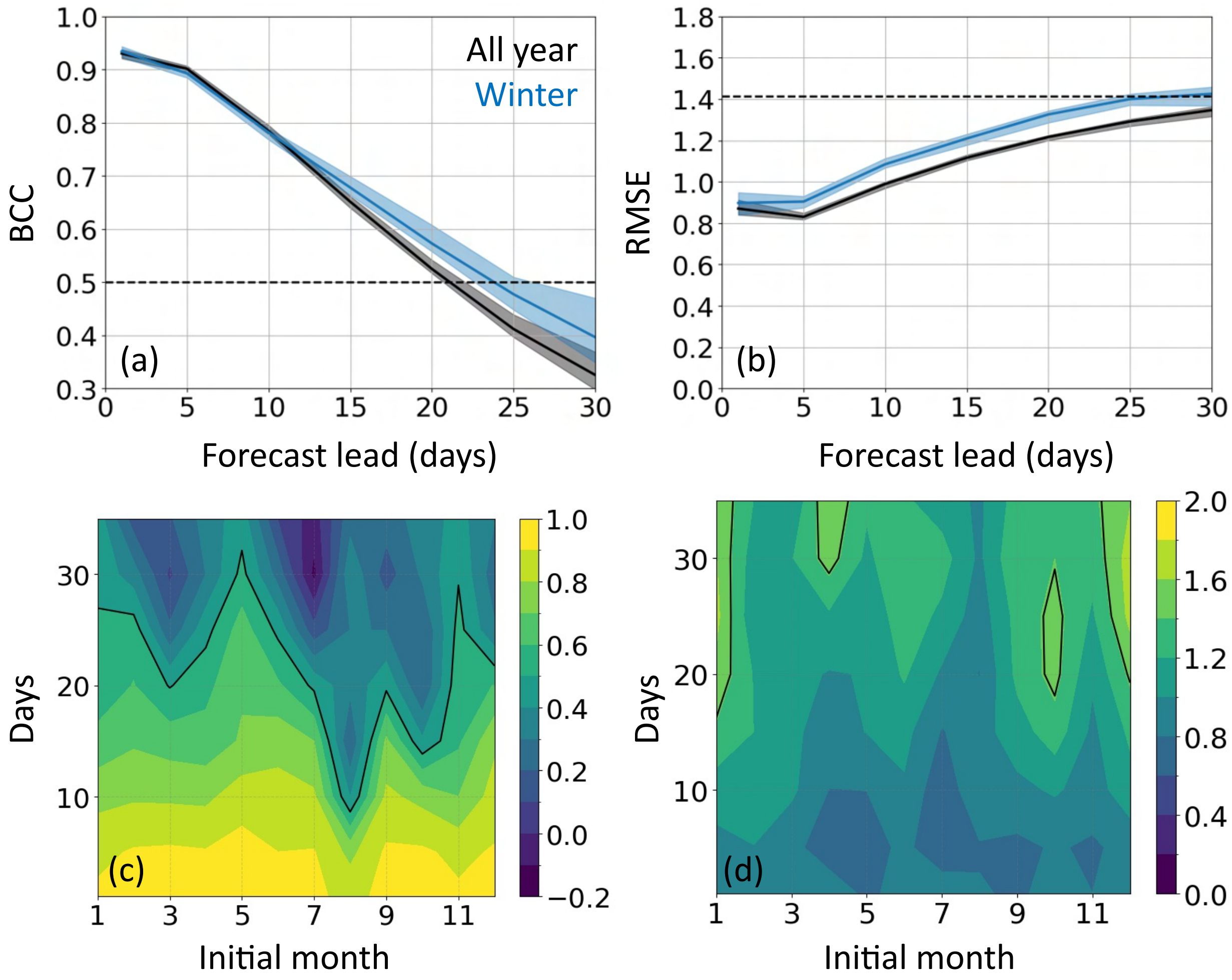

**Figure S10.** Model performance of the standard U-Net shown in Figure S9. (a) BCC for all-year (black) and winter (blue) MJO events whose magnitude is larger than 1. Here, winter refers to November, December, January, and February. The shading indicates model uncertainties from 11 ensemble runs. (b) The corresponding RMSE for all-year and winter MJO events. (c) Sensitivity of BCC to initial months. The result is shown as the average of 11 ensemble runs. (d) Sensitivity of RMSE to initial months. The black lines indicate the thresholds for BCC (0.5) and RMSE ($\sqrt{2}$).

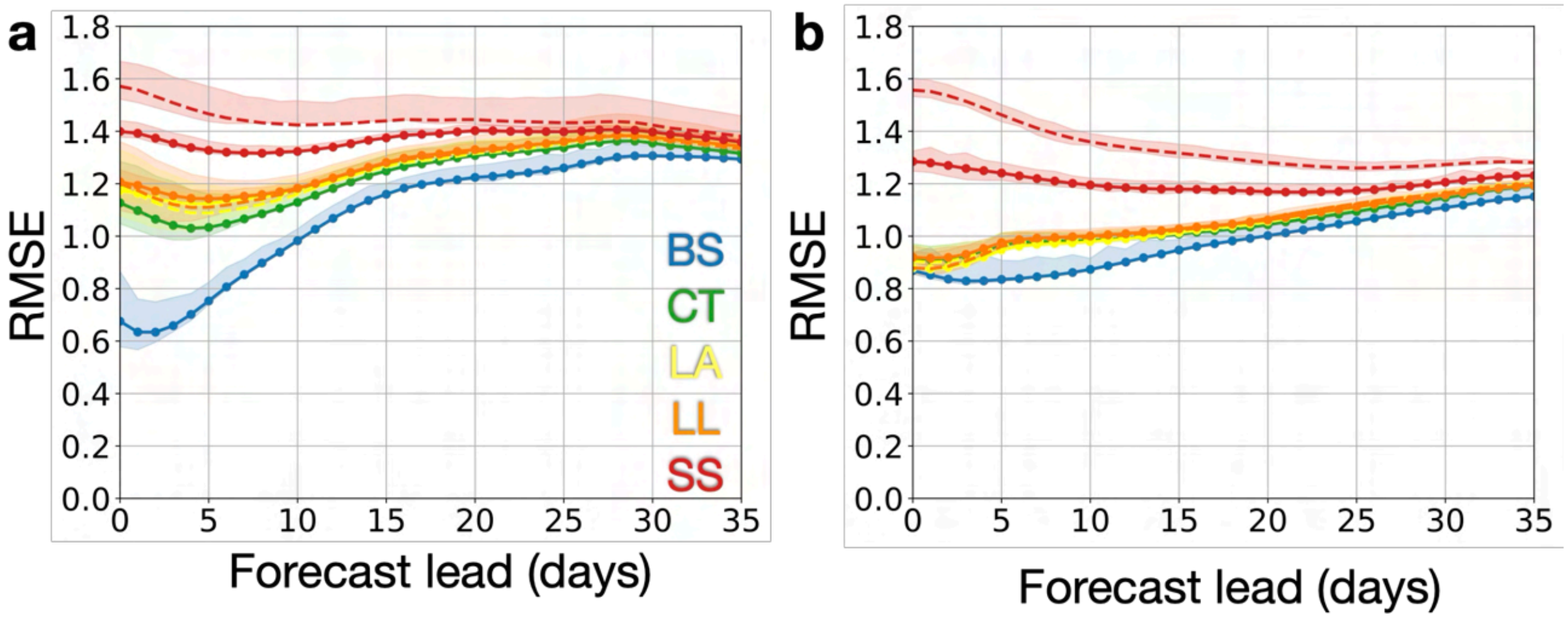

**Figure S11.** Same as Figures 2c-d, but for the root mean squared error (RMSE) of the multi-lead models (DCNN2) in the five experiments (BS-SS).

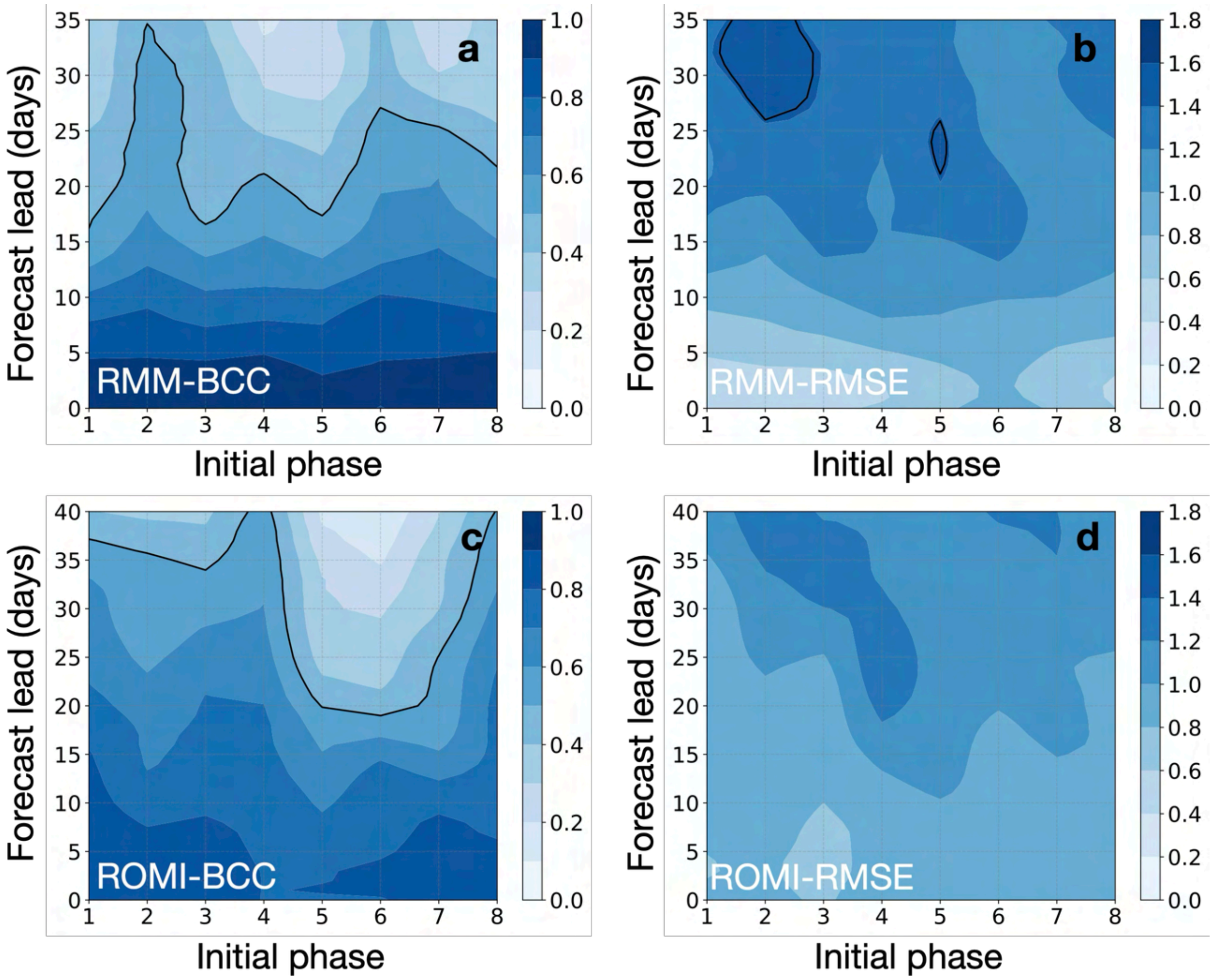

**Figure S12.** Sensitivity of model performance to MJO initial phases for the DCNN2 model using NOAA OLR and 17 additional ERA5 variables as input (BS). The first row is for models predicting RMM, and the second is for models predicting ROMI. The left column shows BCC, and the right column shows RMSE. The black lines indicate the thresholds for BCC (0.5) and RMSE ($\sqrt{2}$). The results are averaged over 100 ensemble runs for all MJO events whose magnitude is larger than 1. These MJO events are divided into eight groups based on their initial phases.

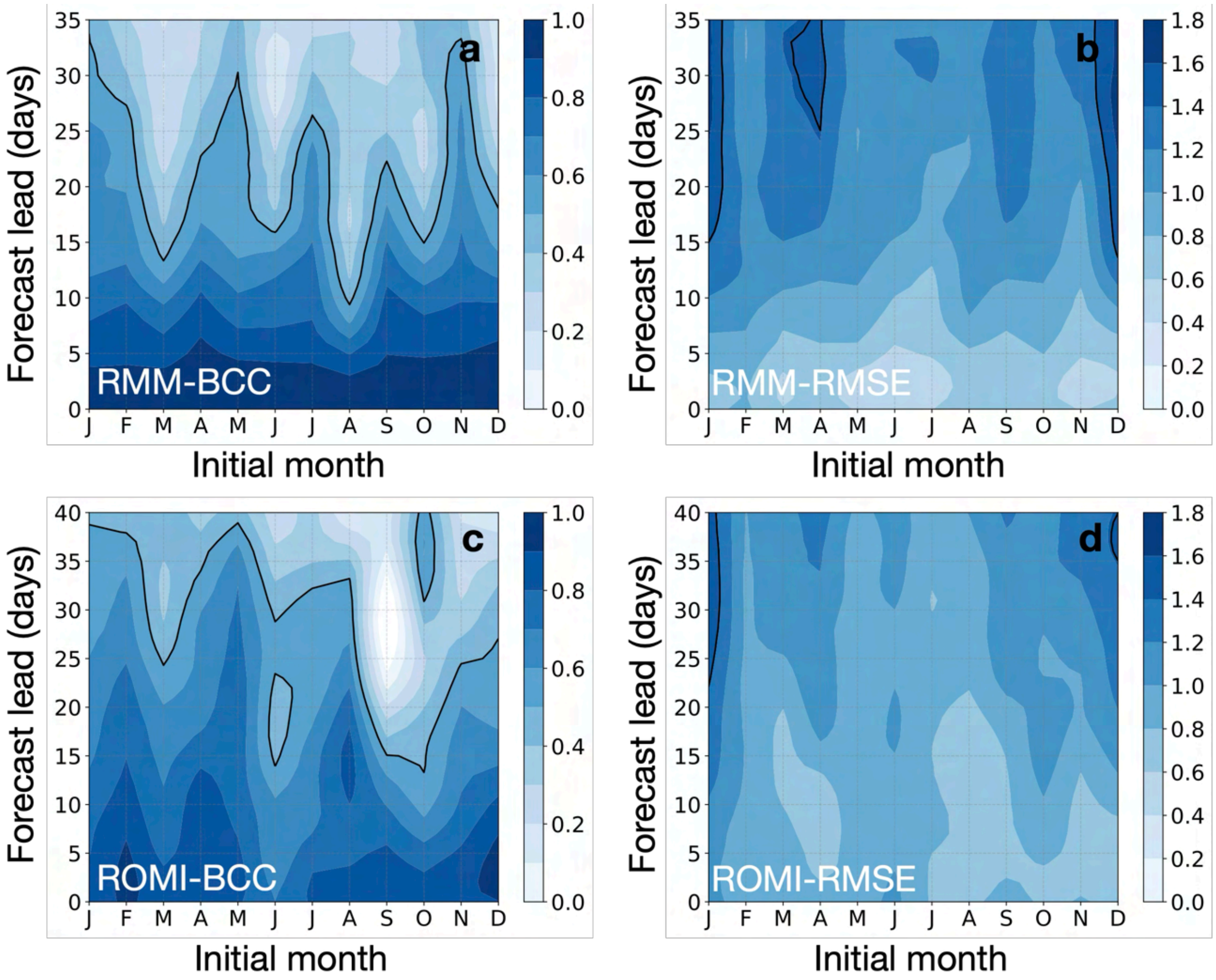

**Figure S13.** Same as Figure S12, but for the sensitivity of model performance to the initial months. Similarly, all MJO events are divided into twelve groups based on their initial months.

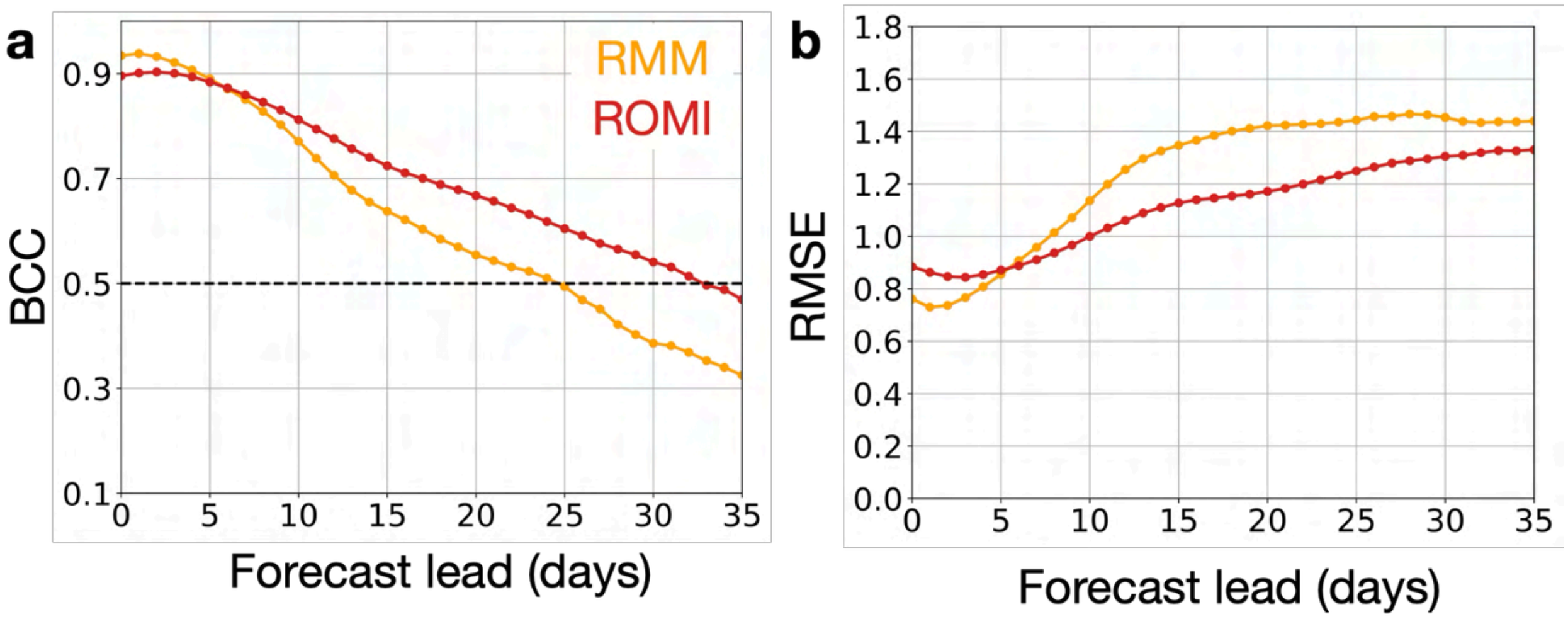

**Figure S14.** Model performance for winter MJO events in December, January, and February. Again, it is DCNN2 model using NOAA OLR and 17 additional ERA5 variables as input (BS).

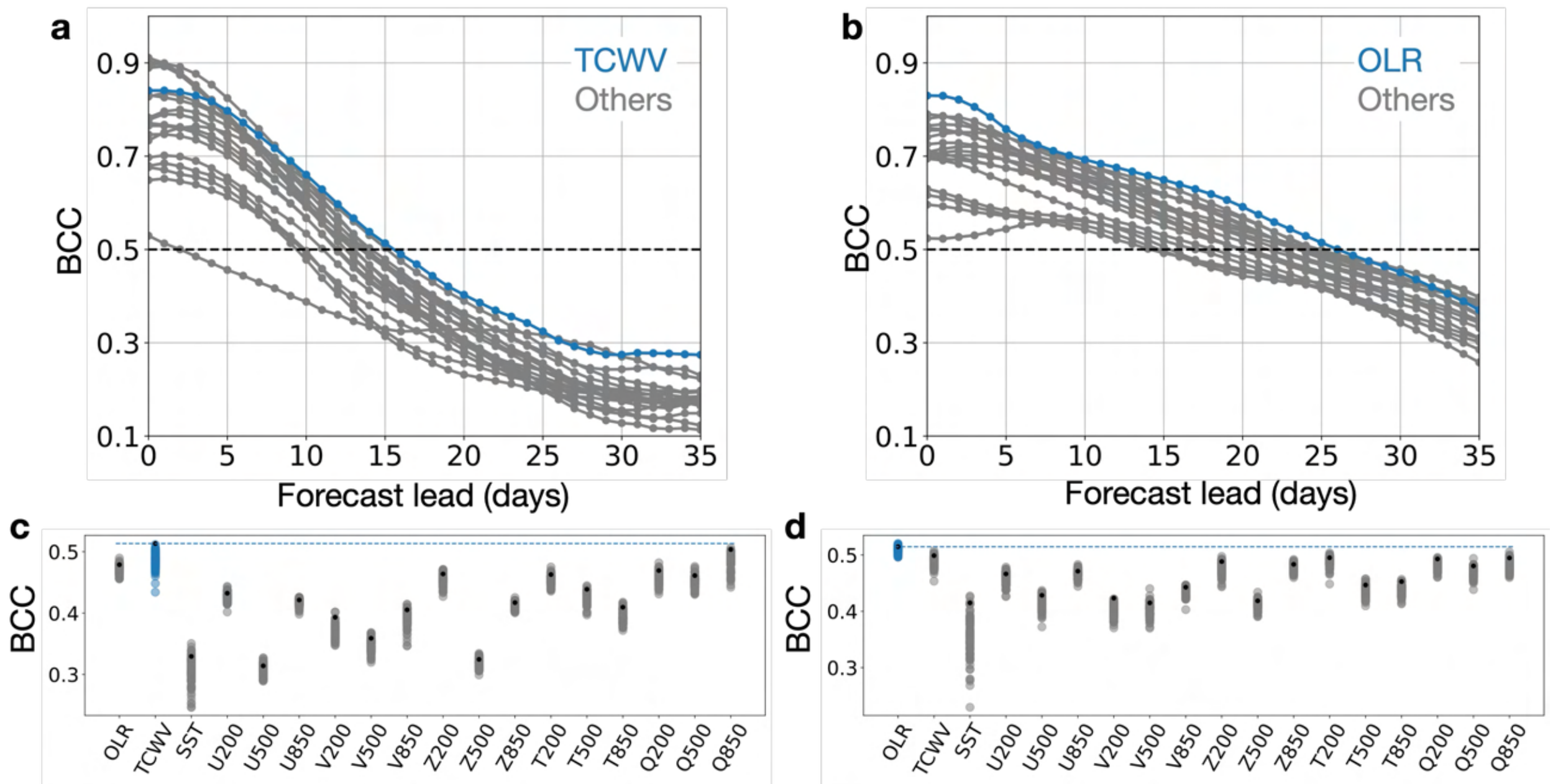

**Figure S15.** BCCs of the multi-lead models using one variable as the input. The left column shows BCC for models predicting RMM, and the right column shows BCC for models predicting ROMI. (a) shows the ensemble-mean BCC as a function of lead times. The blue line is for the model using TCWV as the input, while the grey lines are for models using other variables. (b) is the same as (a) but for models predicting ROMI. (c) shows BCC from eleven individual ensemble runs at lead=15 days for models using different single variables as the input. The spread of grey dots represents model uncertainty. (d) is the same as (c) but for models predicting ROMI at lead=25 days. Note OLR is NOAA OLR.

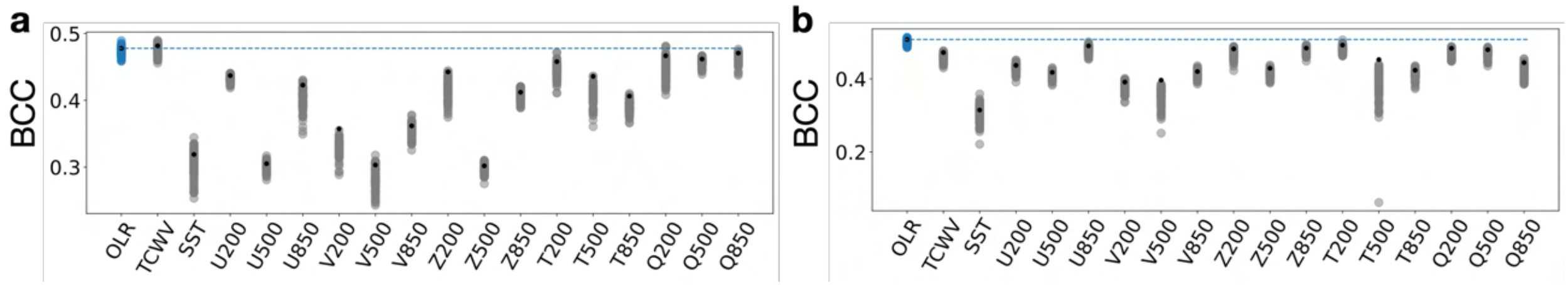

**Figure S16.** Same as Figures S15c-d, but for single-lead models using 18 variables from ERA5.

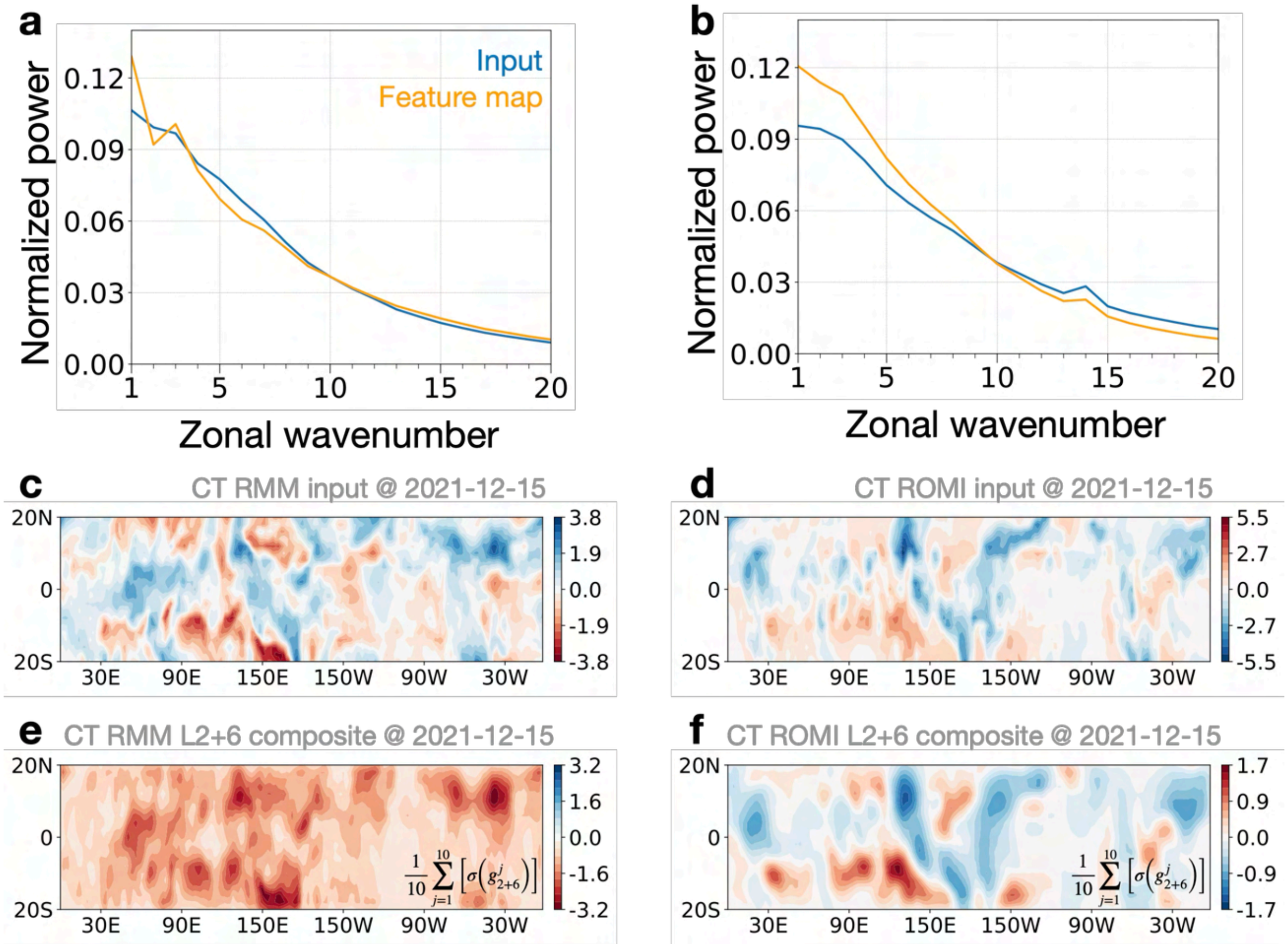

**Figure S17.** Comparison of input fields and feature representations in CT models. Panels (a, b) are analogous to Figure 3, but show Fourier spectra of the inputs and feature maps in the zonal direction only. (a) Results for RMM forecasts using TCWV as input. (b) Results for ROMI forecasts using OLR. Blue lines denote normalized power spectra of the input fields, while orange lines denote spectra of the most influential feature maps from the final convolutional layer (L2+6). Feature maps are selected following the same contribution-based method described in Figure 3. Panels (c–f) are analogous to Figure 4, but for CT models, showing longitude–latitude maps of the inputs (c, d) and composites of the selected L2+6 feature maps (e, f) on 15 December 2021.

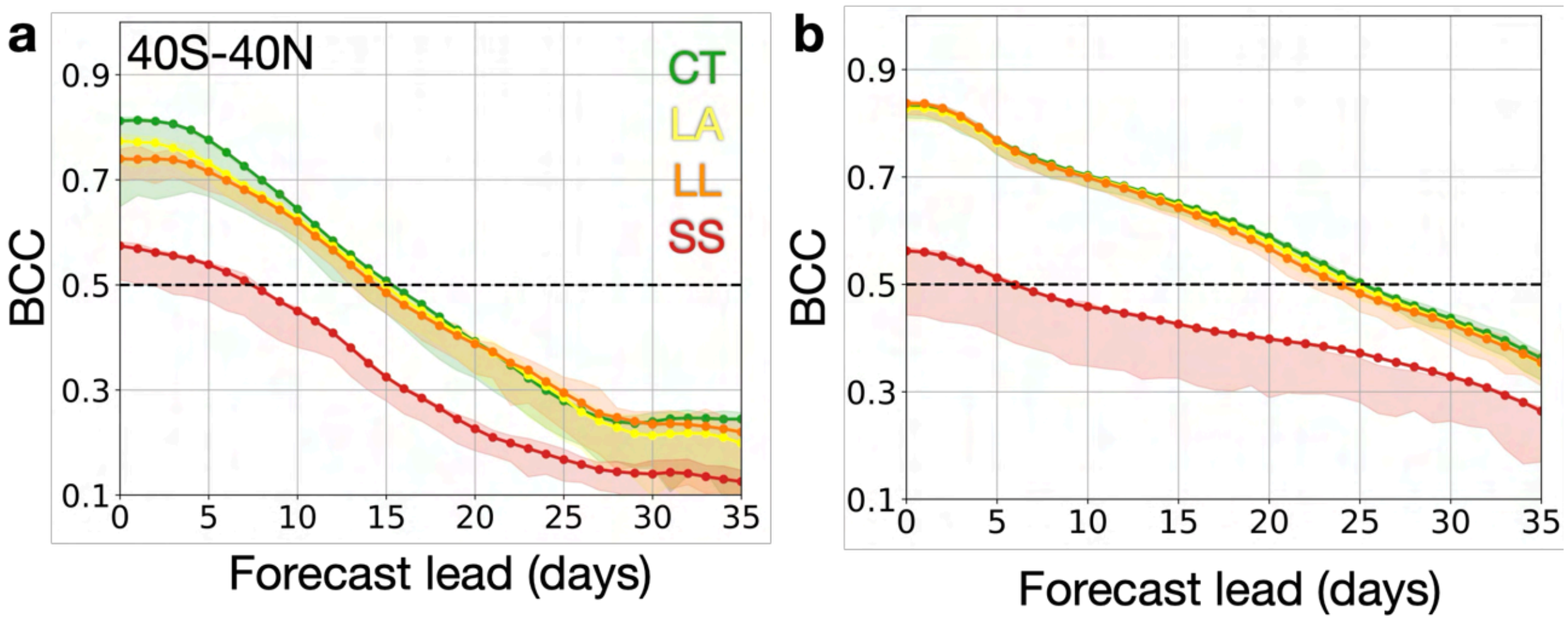

**Figure S18.** Model performance in experiments CT-SS using snapshots over an expanded latitude range (40°S–40°N). Panel (a) shows skill in predicting RMM, and panel (b) shows skill in predicting ROMI. Expanding the meridional domain improves the accuracy of spectral projection in the meridional direction. These experiments confirm the robustness of the sensitivity experiments by addressing potential limitations in the meridional decomposition.

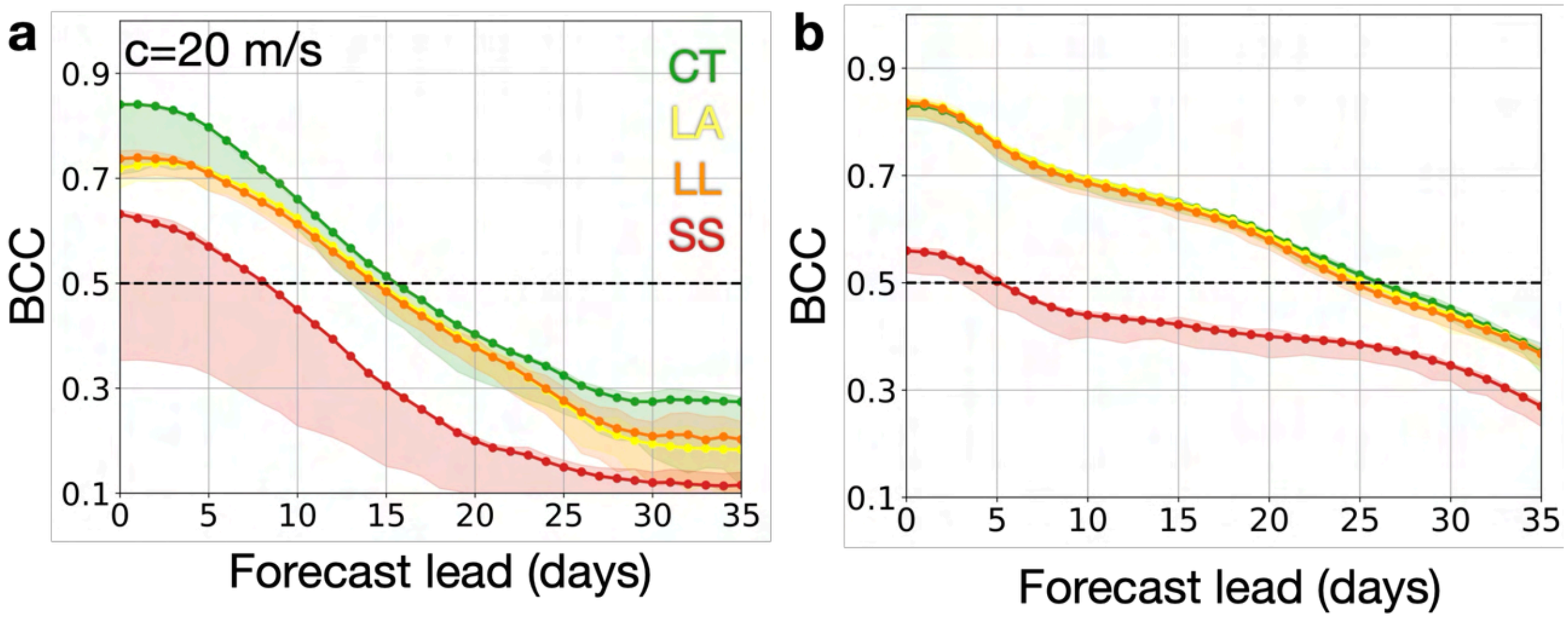

**Figure S19.** Sensitivity of model performance to the gravity wave phase speed used in spectral projection. In the main text, spectral analysis assumes a dry gravity wave speed of c=51 m/s. Here, we repeat the sensitivity experiments (CT-SS) using a reduced wave speed of c=20 m/s, which approximates the propagation speed of moist convectively coupled gravity waves. This choice reflects the influence of moist processes on the deformation radius and spatial scale decomposition. Panel (a) shows RMM prediction skill, and panel (b) shows ROMI prediction skill. The overall conclusion remains unchanged: models using large-scale input (LA and LL) retain most of the skill of the control (CT), while models using only small-scale input (SS) show substantial loss in performance.

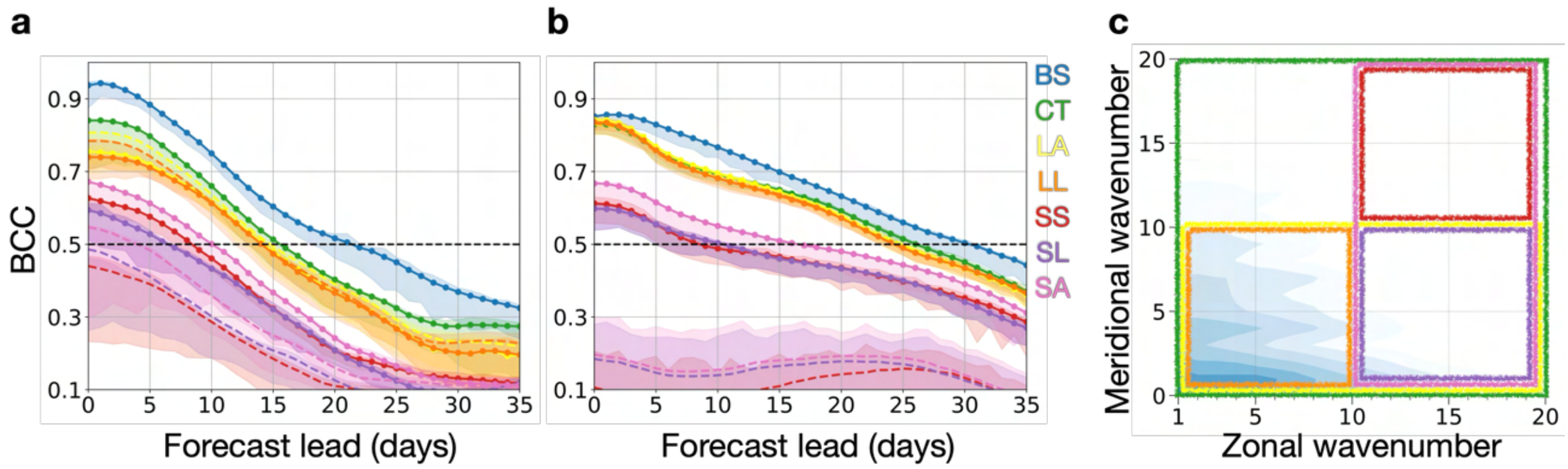

**Figure S20.** Model performance in experiments BS-SA, extending Figures 2c–d by incorporating two additional small-scale experiments (SL and SA). The power spectra in panel (c) indicate the spatial filtering applied to each experiment, with boxes showing the retained wavenumber ranges. Here, SL retains large-scale meridional signals while applying small-scale zonal filtering; SA applies small-scale zonal filtering without meridional filtering. Panel (a) shows the prediction skill for RMM, and panel (b) for ROMI, evaluated using BCC.

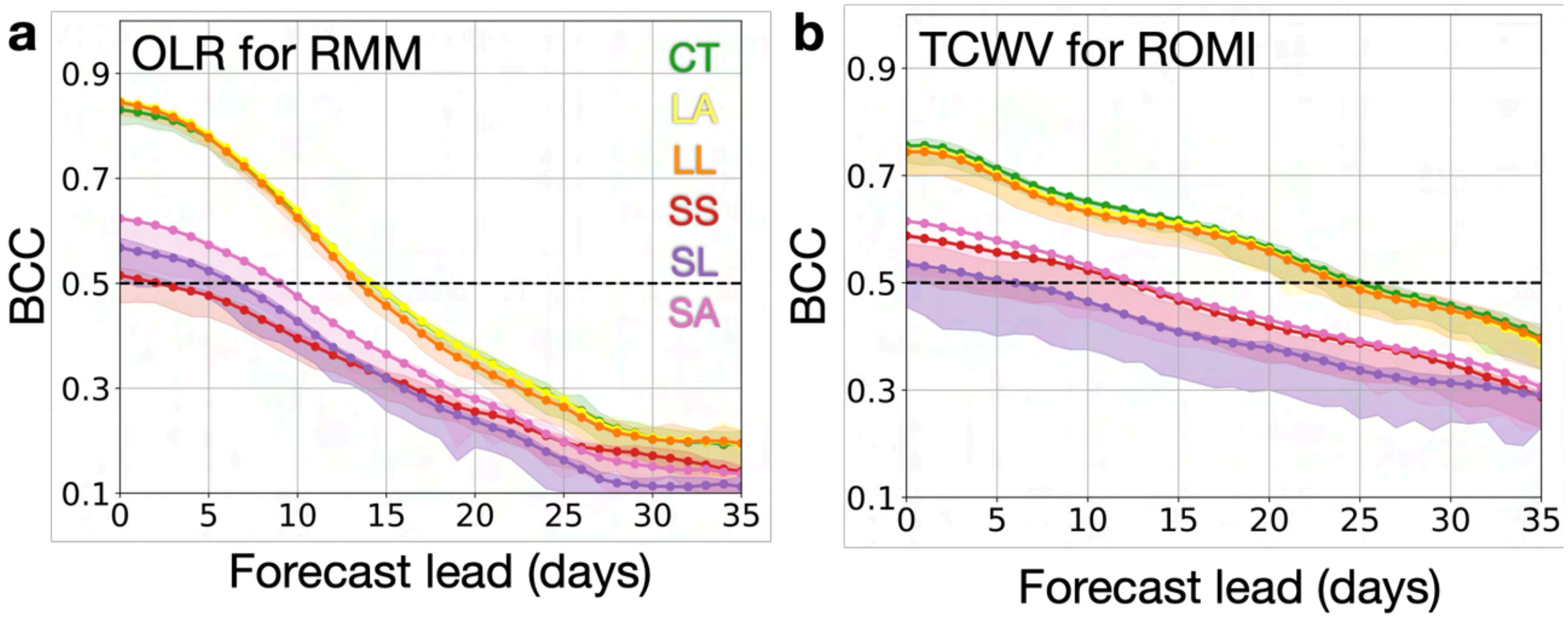

**Figure S21.** Sensitivity of model performance to input variable choice. Panel (a) shows RMM prediction skill using OLR as input, and panel (b) shows ROMI prediction skill using TCWV. The overall conclusions remain robust: large-scale inputs (LA and LL) retain most of the skill from the control experiment (CT), while small-scale experiments (SS, SL, and SA) show significant degradation, regardless of the input variable used.

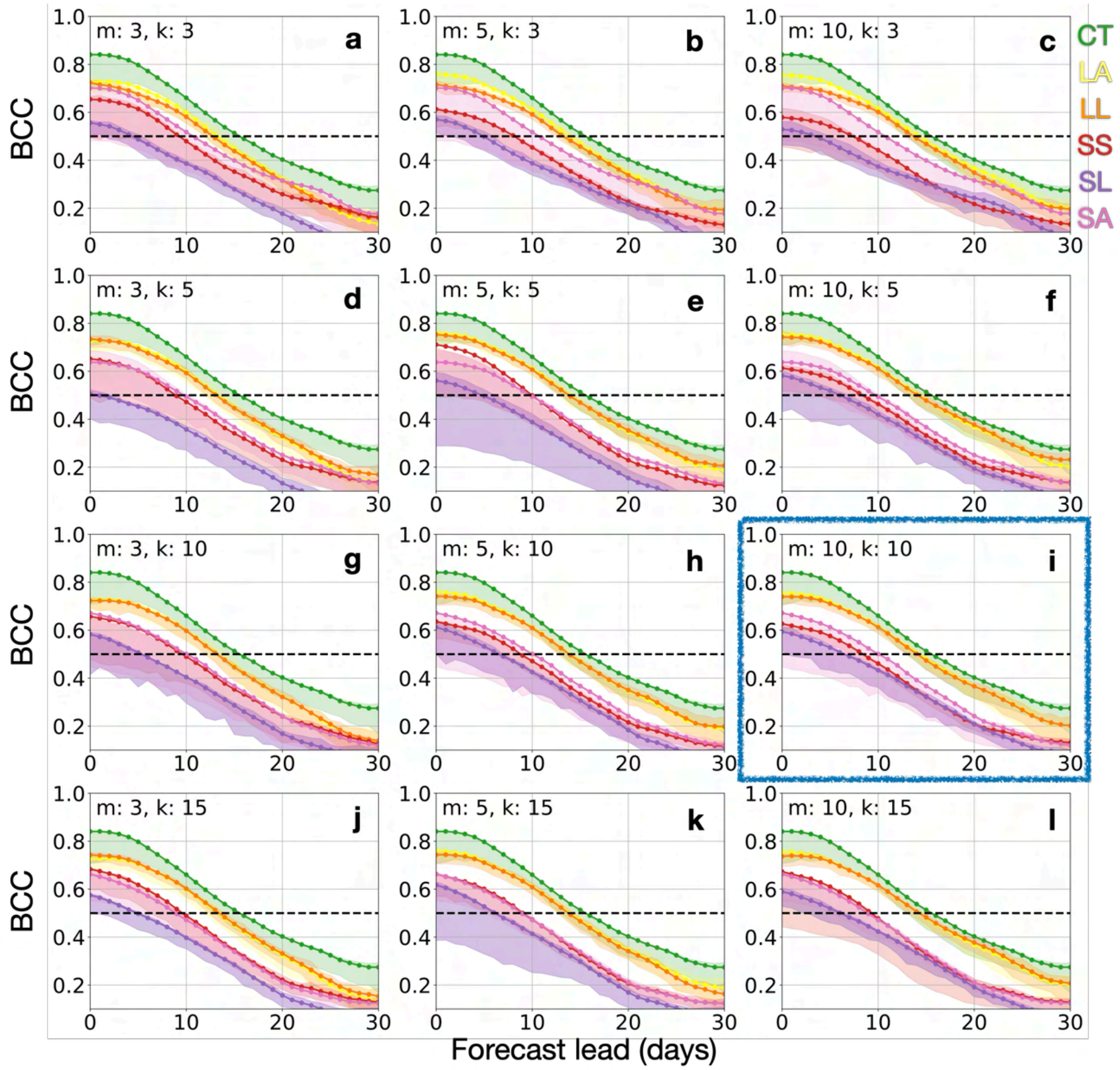

**Figure S22.** Sensitivity test of sensitivity experiments to different cutoff wavenumbers. Results are shown for DCNN2 models predicting RMM. Each panel shows model performance for experiments CT-SA under different combinations of zonal ($k$) and meridional ($m$) cutoff wavenumbers, ranging from $m = 3$, $k = 3$ to $m = 10$, $k = 15$. The default setting is $m = 10$, $k = 10$ (highlighted in blue).

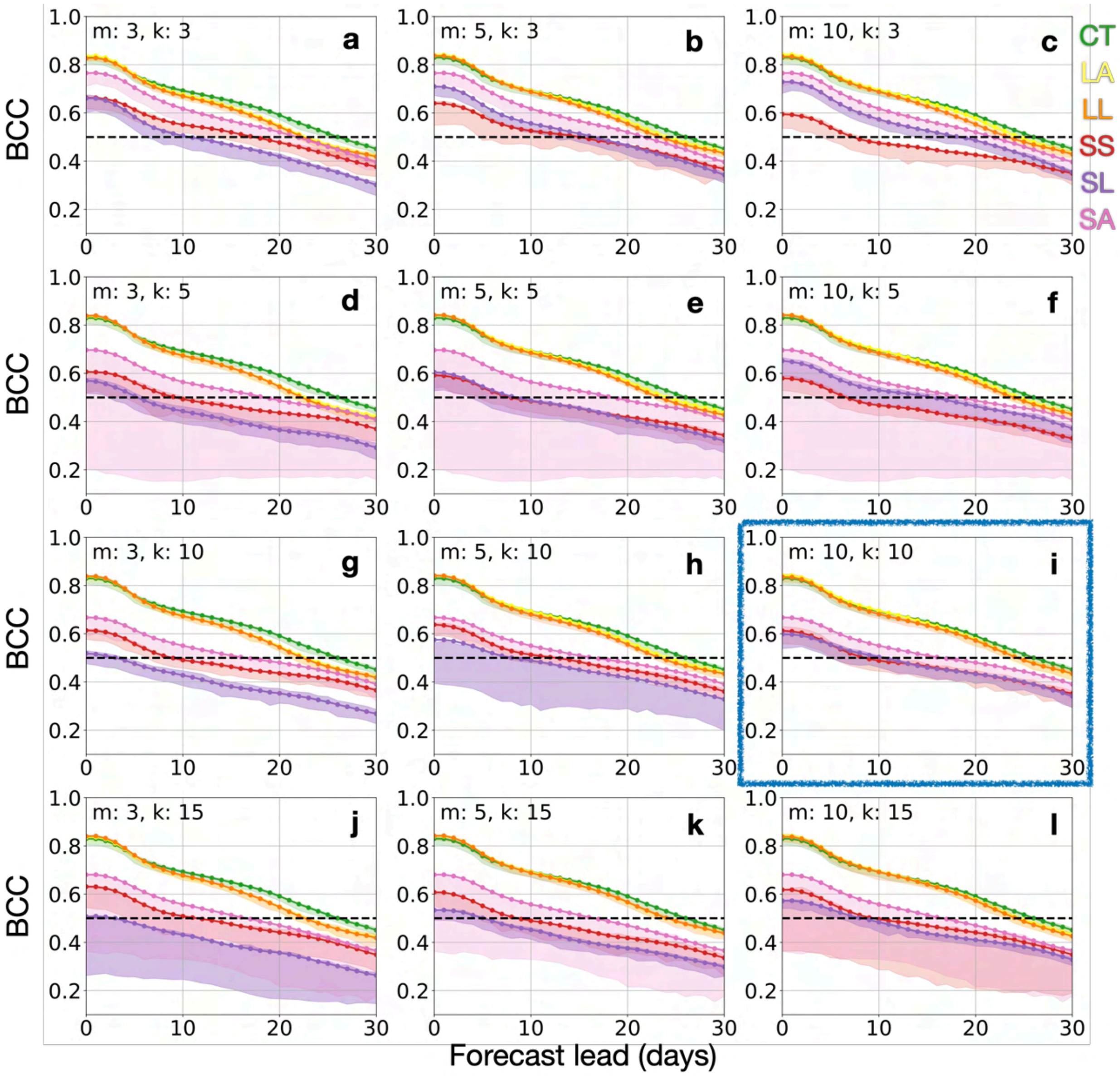

**Figure S23.** Same as Figure S22, but for models predicting ROMI.

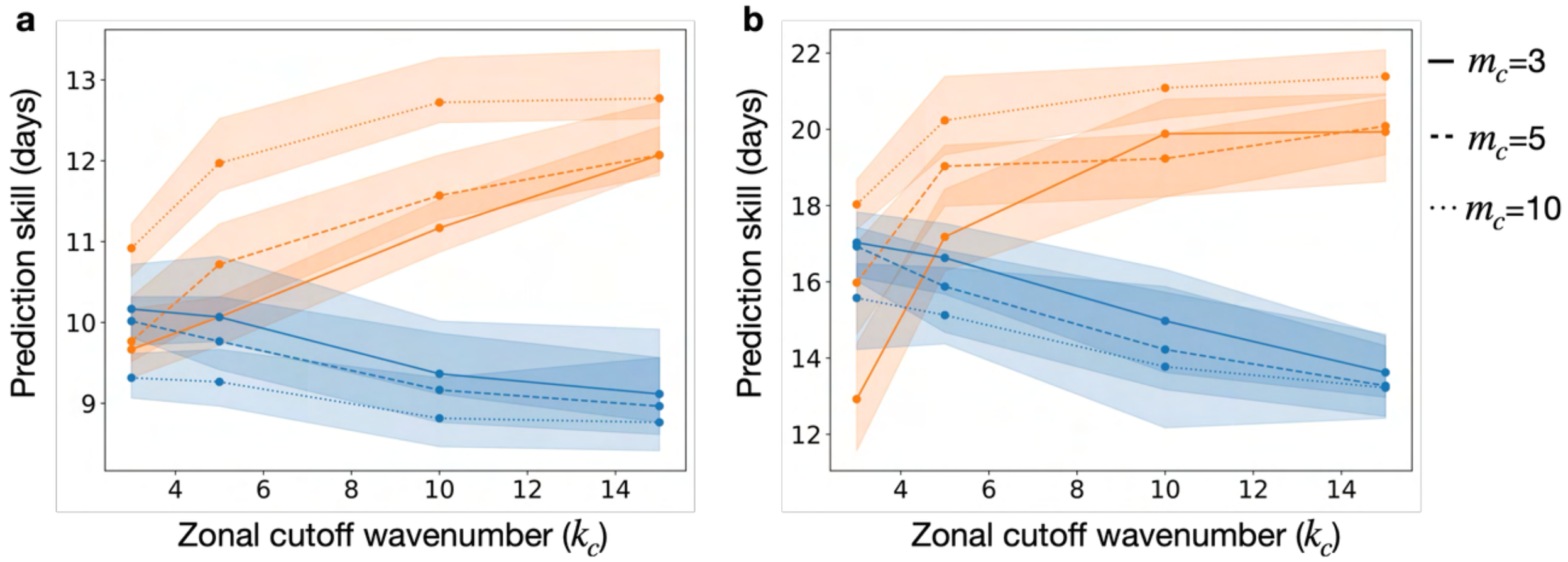

**Figure S24.** Prediction skill as a function of zonal cutoff wavenumber for single-lead models using ERA5 OLR. Panel (a) shows results for RMM, and panel (b) for ROMI. Orange lines represent large-scale experiments (LL), and blue lines represent small-scale experiments (SS). Solid, dashed, and dotted lines indicate meridional cutoff wavenumbers of 3, 5, and 10, respectively. Shading shows uncertainty across 11 ensemble members. Large-scale experiment skill increases with both zonal and meridional cutoff wavenumbers, while small-scale experiment skill decreases. Notably, when the zonal cutoff is set to 3, small-scale experiments outperform large-scale experiments, emphasizing the importance of small-scale processes.

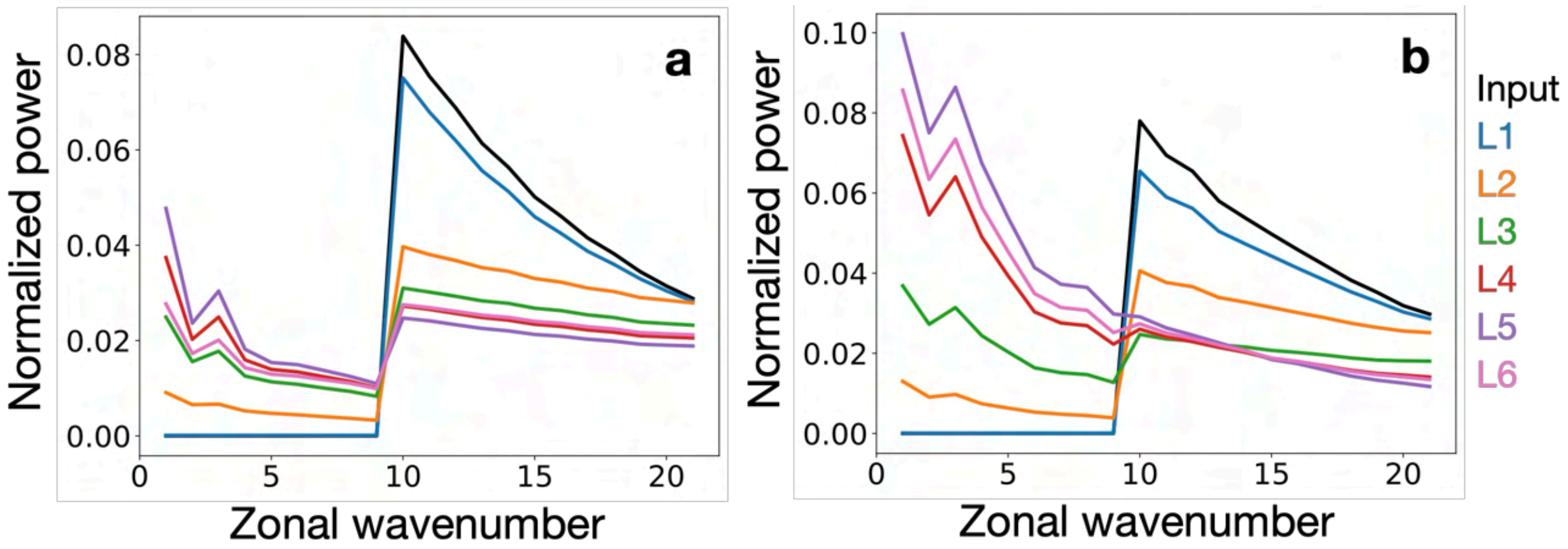

**Figure S25.** Layer-by-layer evolution of zonal-scale power in the small-scale experiment (SS) using multi-lead models. The black line shows the normalized zonal power spectrum of the input, with power at wavenumber zero removed. Colored lines represent the normalized power spectra of the most influential feature maps from layers L1 to L6, before the activation function. Panel (a) shows results for models predicting RMM, and panel (b) for ROMI. The gradual increase in large-scale power across layers indicates that the network progressively reconstructs large-scale structures from small-scale input through convolution and nonlinear activation. Feature map selection follows the same procedure as described in Figure 3. Results from single-lead models are consistent (not shown).

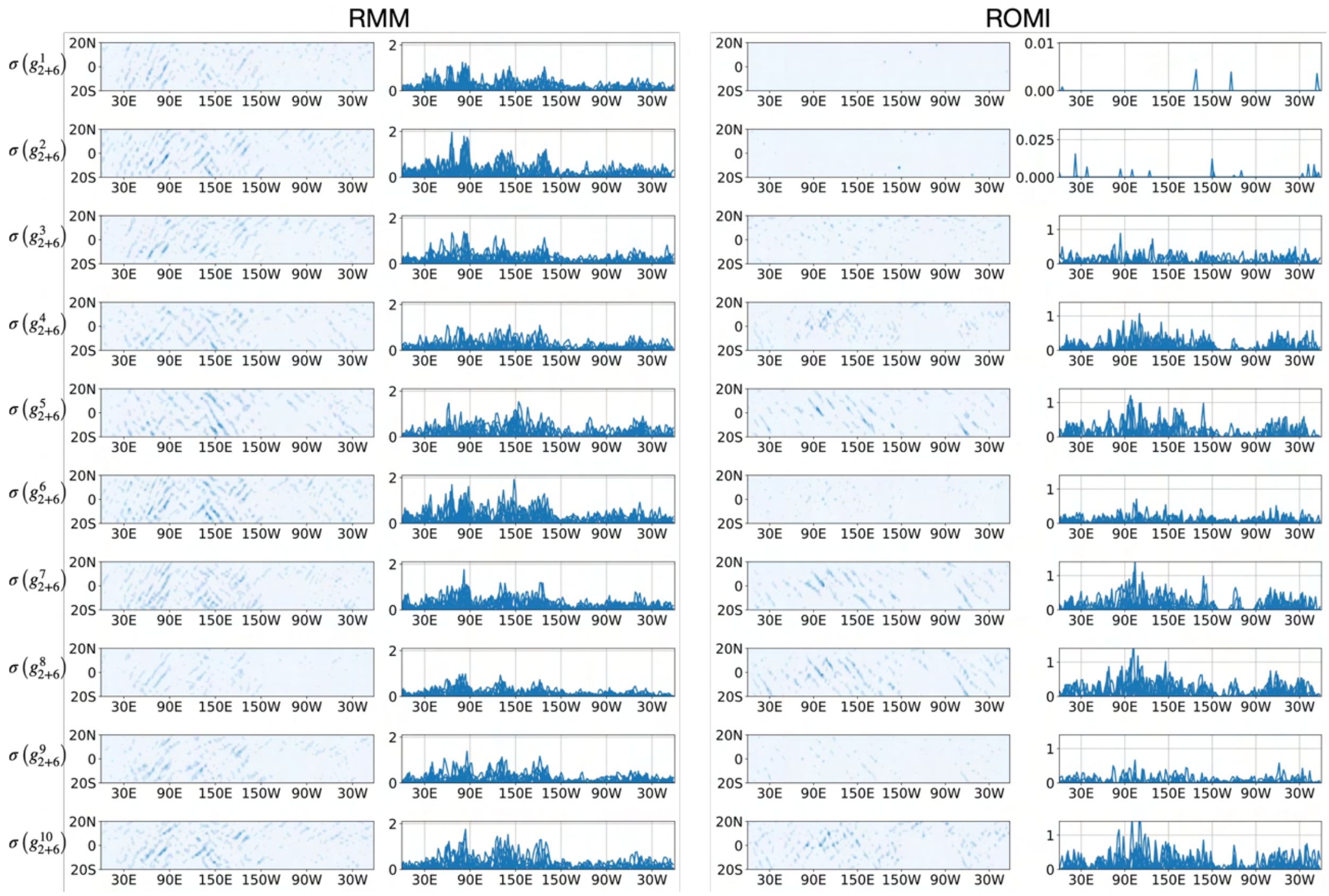

**Figure S26.** Top 10 feature maps from L2+6 that contribute most to the model output, shown for the input on 15 December 2021. The composite maps in Figure 4 are the averages of these 10 feature maps. Most feature maps consistently highlight the large-scale envelopes of the small-scale input signals. An exception occurs in the first two ROMI maps, which display sparse activations that might represent abstract large-scale structures (e.g., convective boundaries).

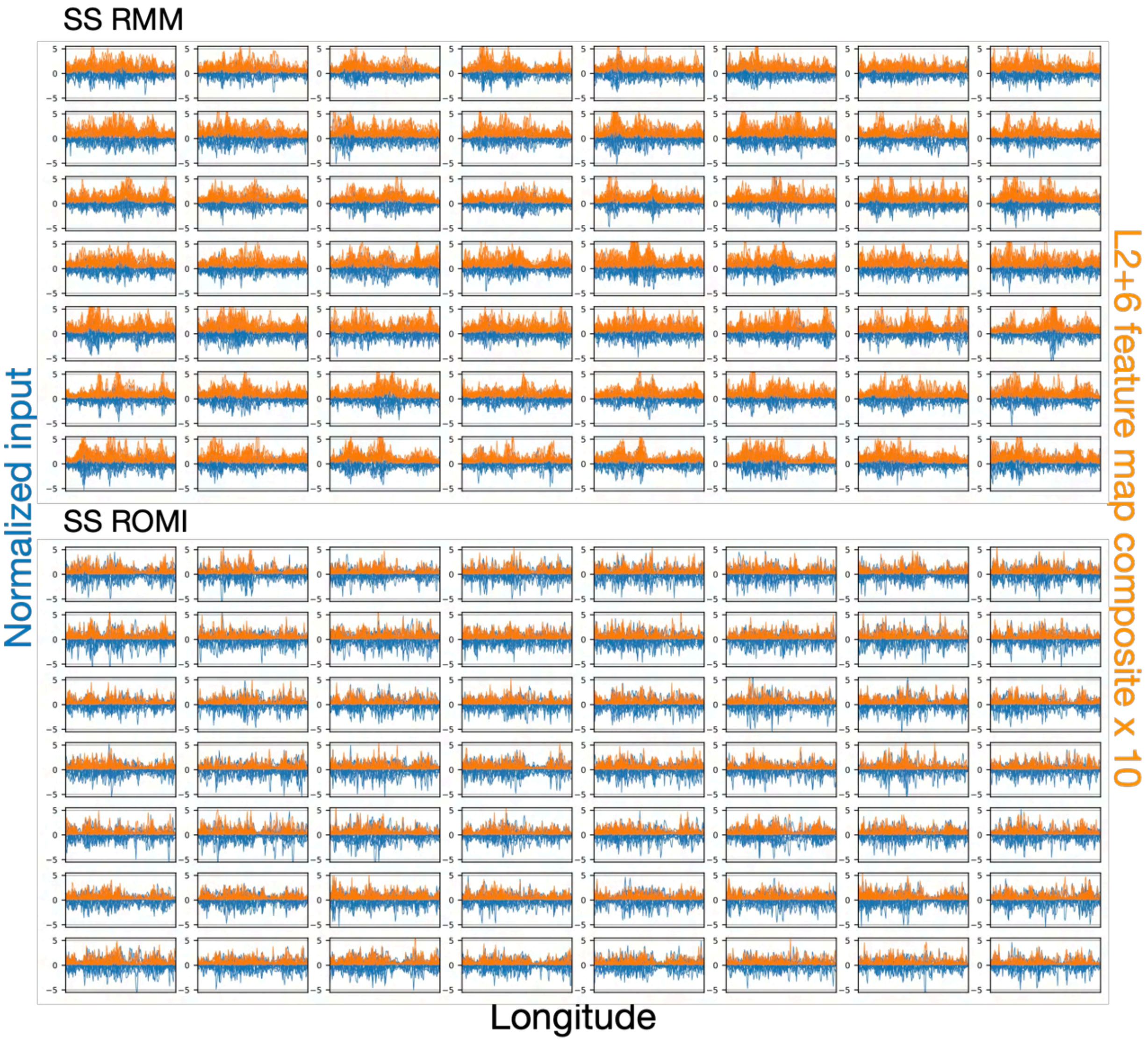

**Figure S27.** Same as Figure 4, but showing a broader set of samples to illustrate the reconstruction of large-scale envelopes from small-scale input in the SS experiment. Blue lines show normalized input fields (zonal profiles at each latitude), while orange lines show the feature map composite from L2+6, amplified by a factor of 10 for visualization. The consistent emergence of large-scale wave packets across the composites highlights the model's ability to recover large-scale structure from small-scale input. Results from single-lead models are consistent (not shown).

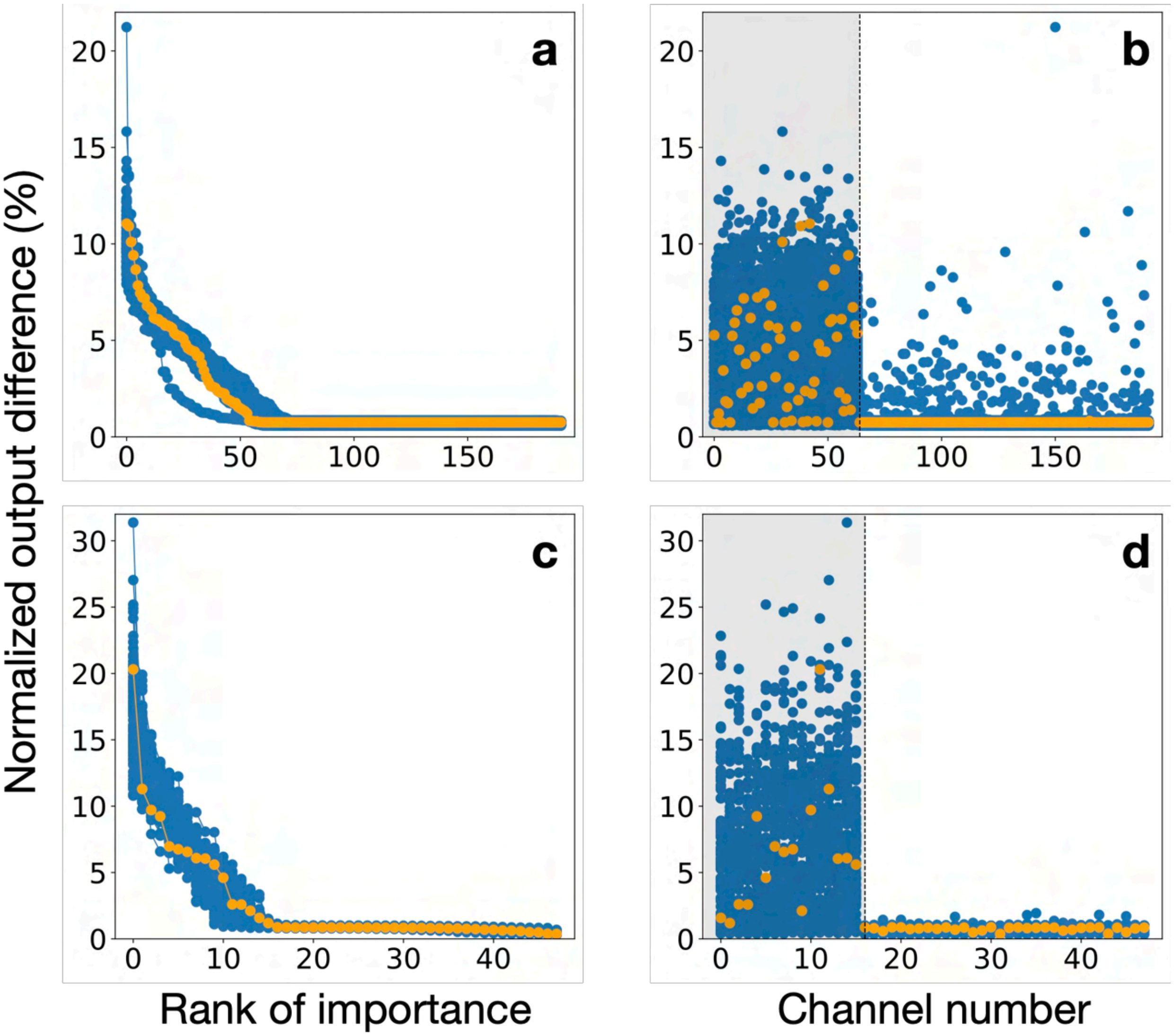

**Figure S28.** Contribution of feature maps from the final convolutional layer (L2+6) to the MJO index predictions. Panels (a–b) show results for RMM, and panels (c–d) for ROMI. Each point represents the normalized output difference after zeroing out a single feature map, expressed as a percentage relative to the original MJO index magnitude. Blue dots represent results from 100 individual ensemble members, while orange dots correspond to the best-performing ensemble member. Panels (a, c) show results sorted by contribution; panels (b, d) show unsorted results, with feature maps from L2 highlighted in gray. This analysis quantifies the influence of individual feature maps on model output and follows the approach of Shin et al. (2022), with the key modification that entire feature maps are zeroed rather than localized regions.

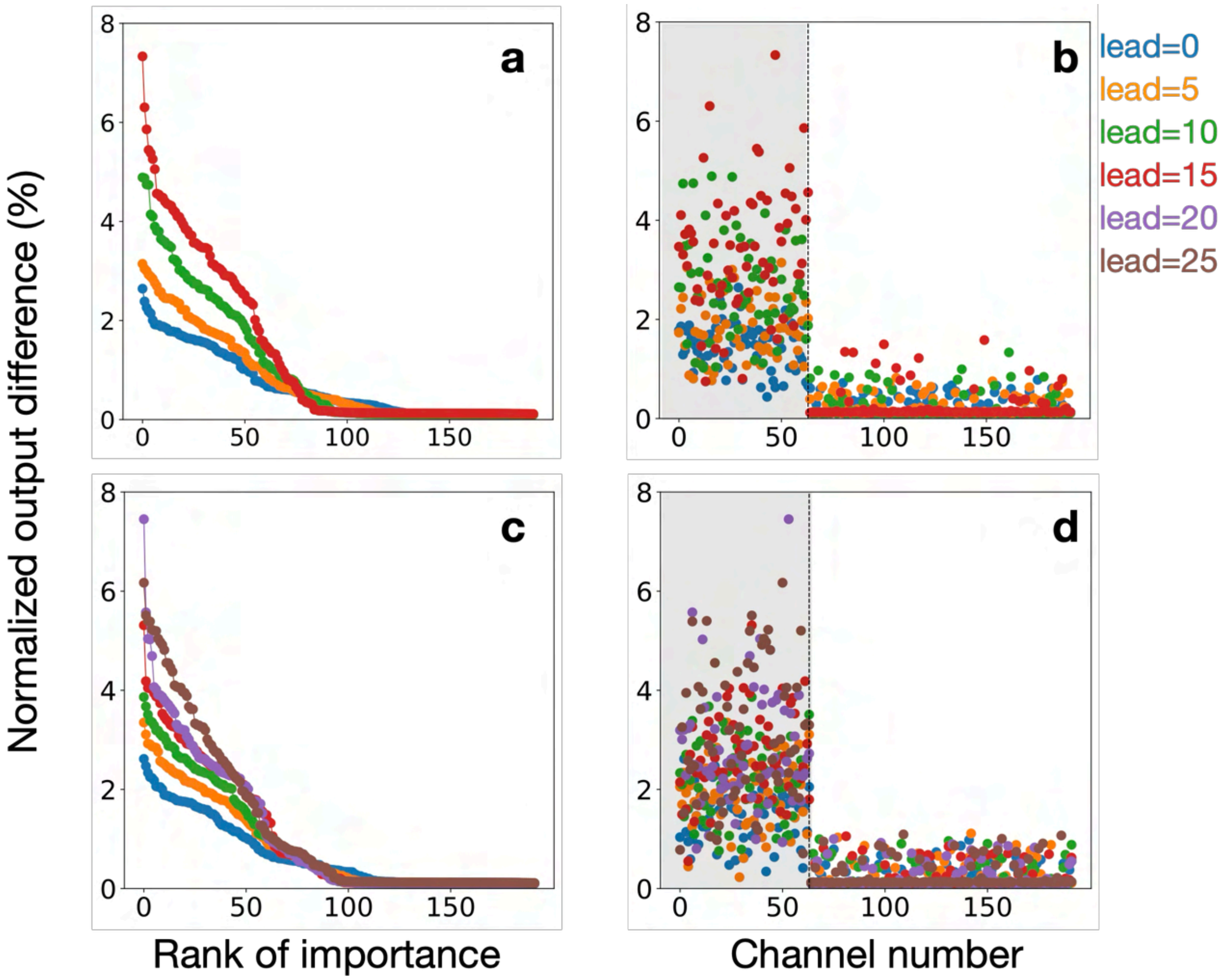

**Figure S29.** Same as Figure S28, but for single-lead models using ERA5 OLR. Results shown are from the first ensemble member only. Different colors indicate models at different lead times. This analysis confirms the consistent importance of L2 feature maps across forecast lead times.

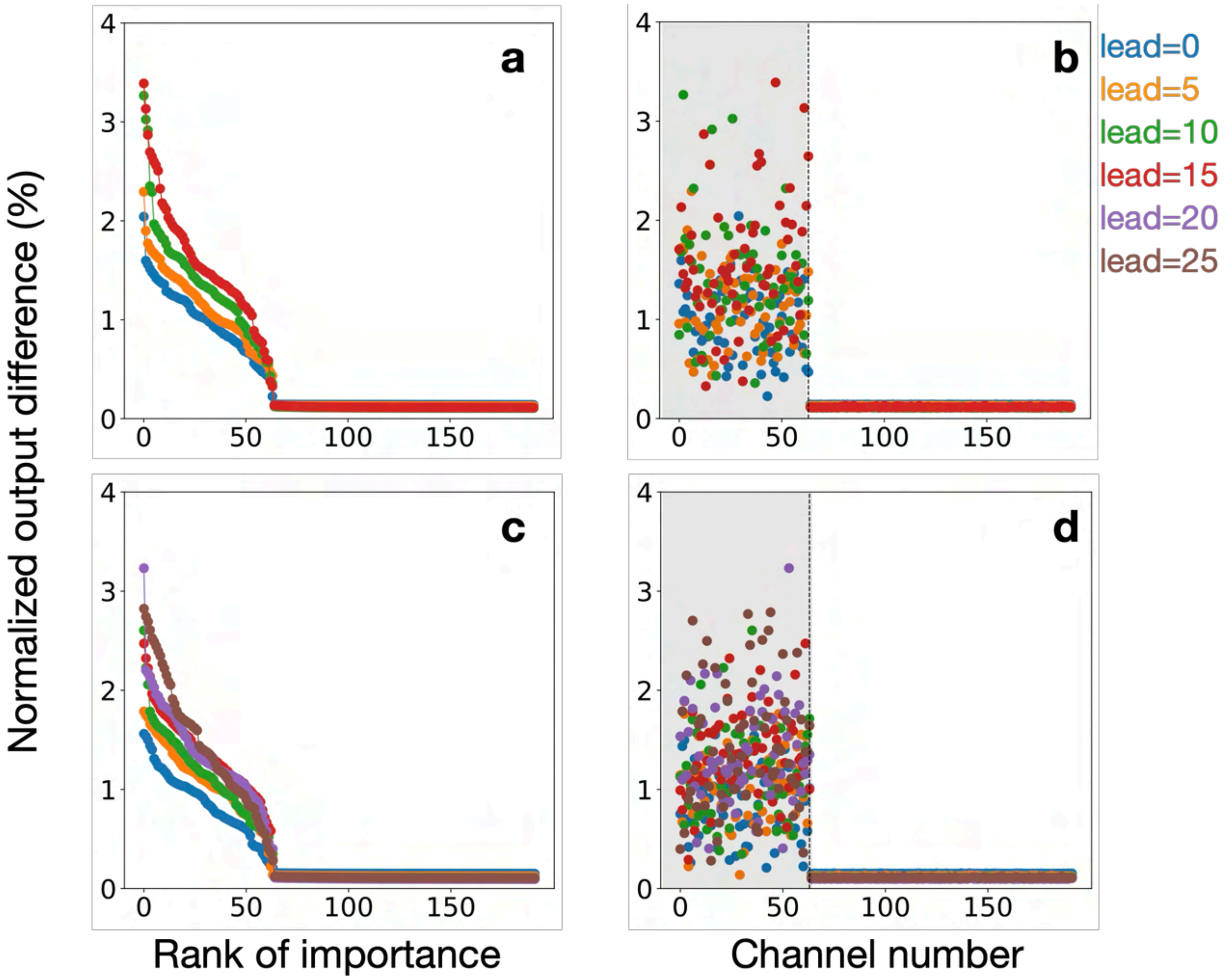

**Figure S30.** Same as Figure S29, but feature map contributions are evaluated by adding small perturbations rather than zeroing out. Results from this perturbation-based method are consistent with those obtained through zeroing (Figure S29), reinforcing the dominant role of shallower layers in MJO prediction.

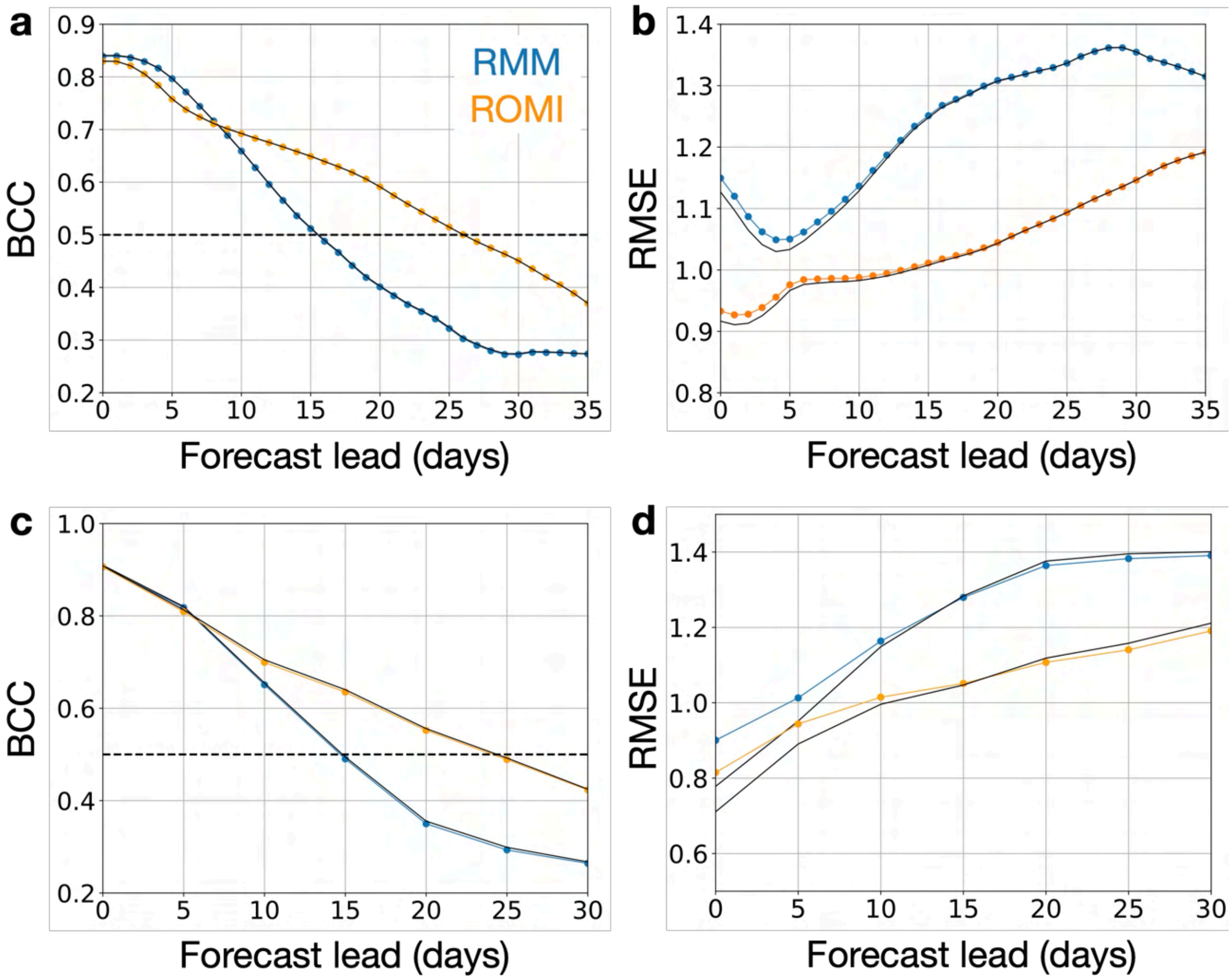

**Figure S31.** Model performance for distilled CT models in which all feature maps from layers L3 to L6 are zeroed out post-training, without retraining the model. Top row: results for multi-lead models using NOAA OLR. Bottom row: results for single-lead models using ERA5 OLR. Colored lines represent ensemble-mean performance of the distilled models, while black lines show the corresponding ensemble means for the original (full) CT models. The close alignment between the two curves indicates that deeper convolutional layers contribute little to MJO prediction skill.

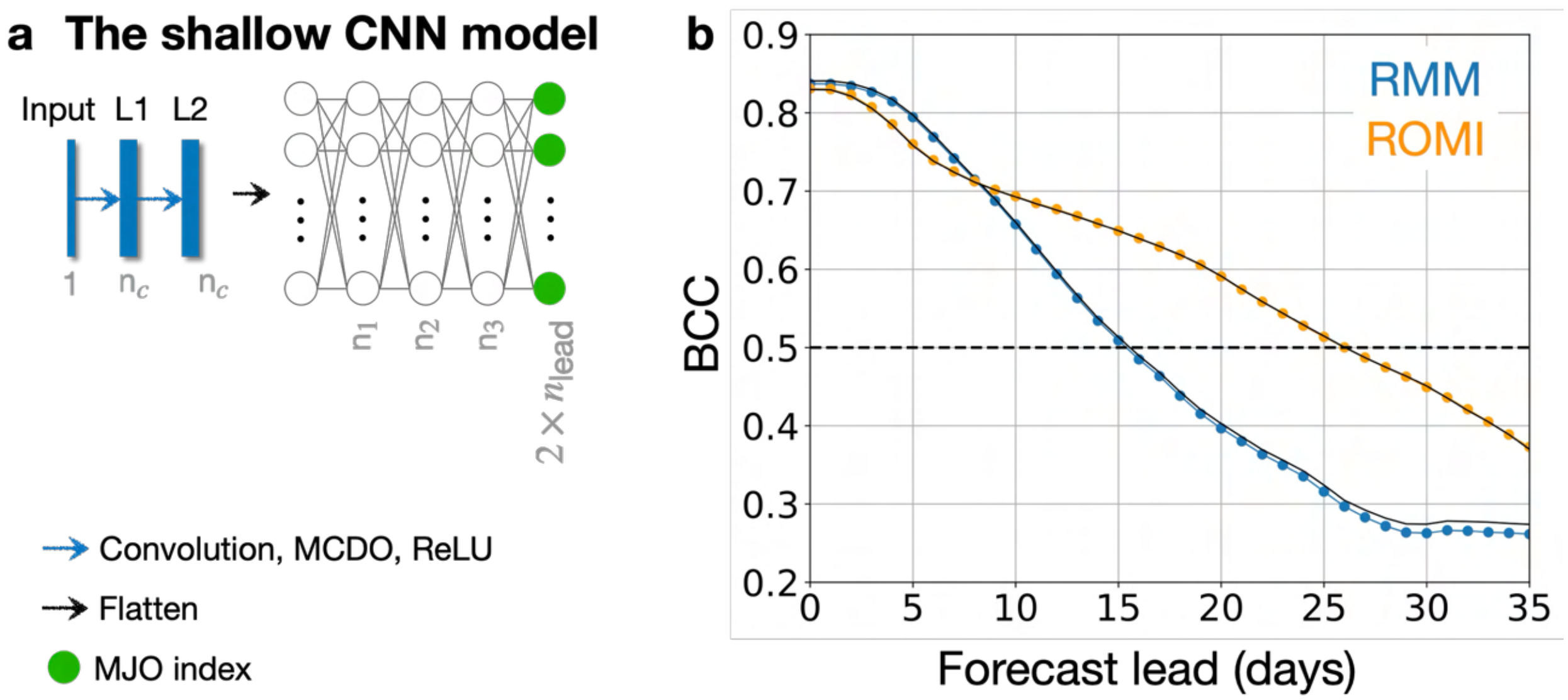

**Figure S32.** Shallow CNN architecture and performance comparison with the original DCNN model. (a) Architecture of the shallow CNN model, which retains the same setup as the DCNN shown in Figure 2b but reduces the depth to include only the first two convolutional layers (L1 and L2). This reduction lowers the number of trainable parameters in CT models from 436 million to 145 million for RMM prediction, and from 73 million to 24 million for ROMI prediction. (b) Model performance of the shallow CNN. Colored lines indicate ensemble-mean skill across multiple lead times, while black lines show the corresponding ensemble means from the full DCNN CT models in Figure 2b. The comparable performance suggests that deeper convolutional layers (L3–L6) contribute minimally to MJO forecasting skill.

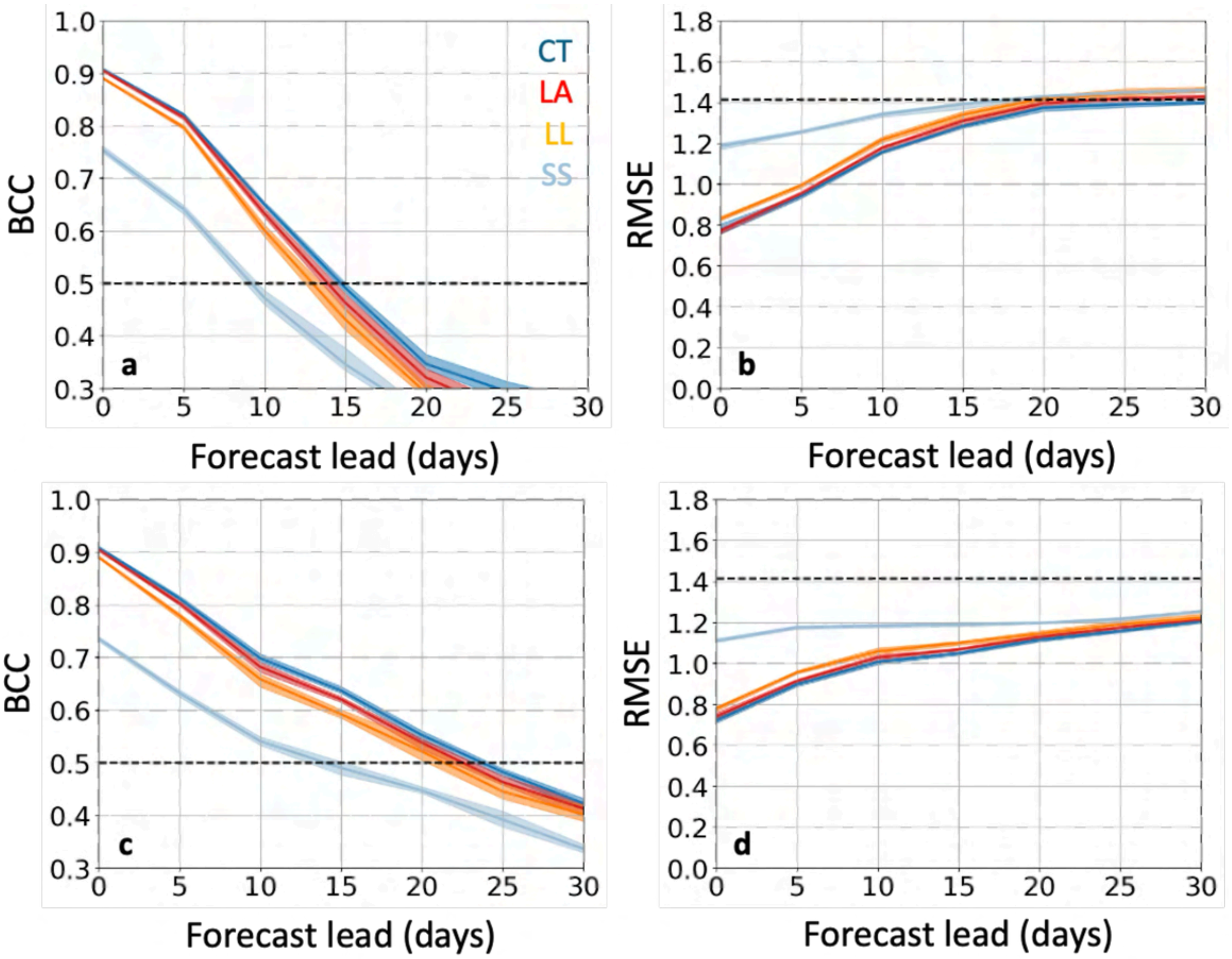

**Figure S33.** Model performance across experiments CT-SS using the single-lead shallow CNN based on ERA5 OLR (architecture in Figure S32a). Models in LA-SS are retrained from scratch with the corresponding filtered data. The top row shows prediction skill for RMM, and the bottom row shows results for ROMI. Results from the shallow CNN are consistent with those from the deeper DCNN models, confirming the robustness of findings in the main text. Similar filtered-input experiments can be applied to other architectures.

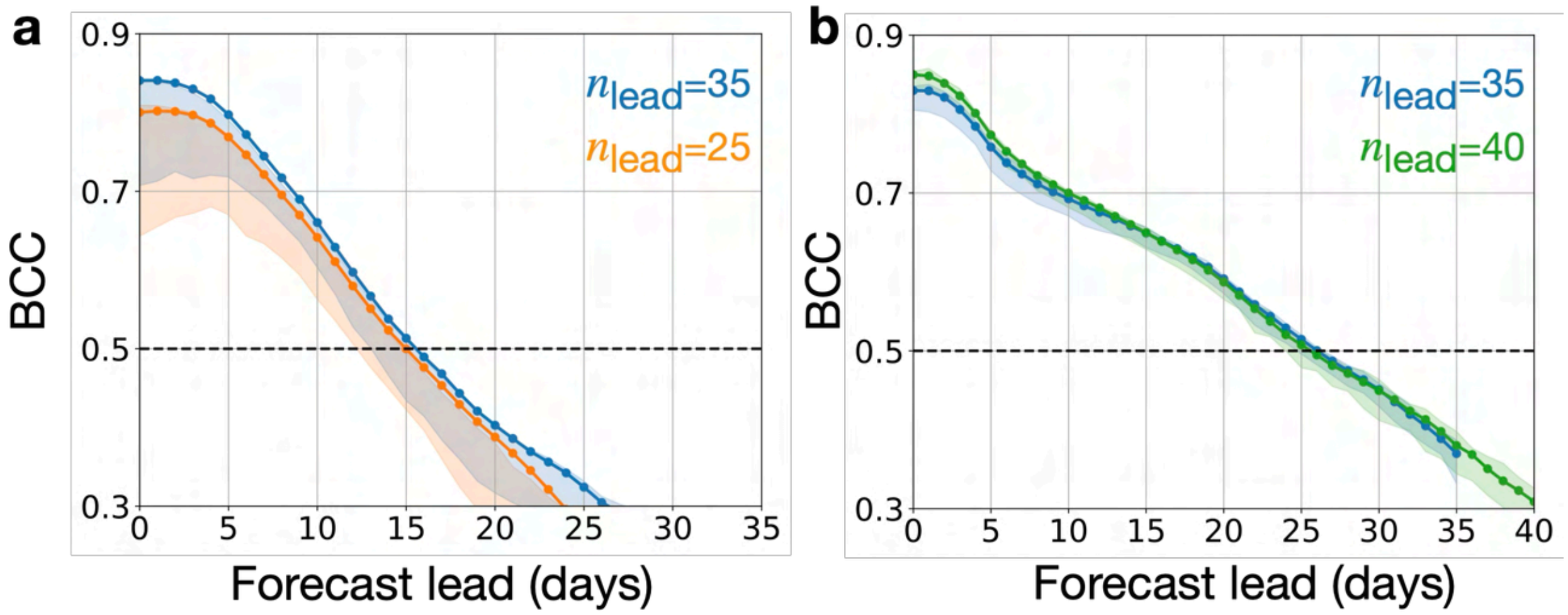

**Figure S34.** Sensitivity of multi-lead model performance to the number of prediction lead times ($n_{\text{lead}}$). (a) Comparison between $n_{\text{lead}} = 25$ and $n_{\text{lead}} = 35$ for models predicting RMM. (b) Same as (a), but for models predicting ROMI. Results indicate that model performance is not sensitive to moderate changes in the number of lead times used in training.

| Name | BS | CT | LA | LL | SS |
|---|---|---|---|---|---|
| Input number $n$ | 18 | 1 | 1 | 1 | 1 |
| Kernel number $n_c$ | 32 | 16 | 16 | 48 | 32 |
| $1^{st}$ FCN nodes $n_1$ | 500 | 400 | 200 | 500 | 600 |
| $2^{nd}$ FCN nodes $n_2$ | 150 | 150 | 250 | 200 | 150 |
| $3^{rd}$ FCN nodes $n_3$ | 70 | 100 | 50 | 80 | 80 |
| Learning rate | 0.000227 | 0.001556 | 0.004457 | 0.000252 | 0.000100 |
| Batch size | 90 | 30 | 100 | 20 | 70 |
| Dropout rate | 0.126477 | 0.166492 | 0.251792 | 0.399000 | 0.322332 |
| Kernel size | 3x3 | 5x5 | 5x5 | 7x7 | 3x3 |
| Optimizer | Adam | SGD | SGD | SGD | Adam |
| Size | 182 | 73 | 36 | 273 | 218 |

**Table S1.** Best hyperparameters for DCNN2 models predicting ROMI using NOAA OLR across experiments BS-SS. Listed parameters include input channel number ($n$), number of convolutional kernels ($n_c$), number of nodes in the first three fully connected layers ($n_1$, $n_2$, $n_3$), learning rate, batch size, dropout rate, kernel size, optimizer, and the number of trainable parameters in millions. Hyperparameters for other models (e.g., those predicting RMM or using alternate configurations) are available upon request.